\documentclass{pasa}
\usepackage{color}
\usepackage{graphicx}
\usepackage{amsmath}
\usepackage{txfonts}
\usepackage{aas_macros}
\usepackage{hyperref}
\hypersetup{colorlinks, citecolor=blue, linkcolor=blue, urlcolor=blue}

\title[Visualising the Kinematics of Planet Formation]{Visualising the Kinematics of Planet Formation}

\author{{\bf Disk Dynamics Collaboration$^{\star}$},                        
Philip J. Armitage$^{1,2}$,                                                 
Jaehan Bae$^{3}$\thanks{NASA Hubble Fellowship Program Sagan Fellow},       
Myriam Benisty$^{4}$,                                                       
Edwin A. Bergin$^{5}$,                                                      
Simon Casassus$^{6}$                                                        
Ian Czekala$^{7*}$                                                          
Stefano Facchini$^{8}$,                                                     
Jeffrey Fung$^{9}$,                                                         
Cassandra Hall$^{10,11,12}$,                                                
John D.\ Ilee$^{13}$,                                                       
Miriam Keppler$^{14}$,                                                      
Aleksandra Kuznetsova$^{5}$,                                                
Romane Le Gal$^{15}$,                                                       
Ryan A. Loomis$^{16}$,                                                      
Wladimir Lyra$^{17}$,                                                       
Natascha Manger$^{1}$,                                                      
Sebasti\'an P\'erez$^{18}$,                                                 
Christophe Pinte$^{4,19}$,                                                  
Daniel J. Price$^{19}$,                                                     
Giovanni Rosotti$^{20}$,                                                    
Judit Szul\'{a}gyi$^{21}$,                                                  
Kamber Schwarz$^{22*}$,                                                     
Jacob B. Simon$^{23}$,                                                      
Richard Teague$^{15,\star}$,                                                
Ke Zhang$^{5,24}$                                                           
\affil{$^{\star}$Corresponding Author: \url{richard.d.teague@cfa.harvard.edu}}\vspace{0.3cm} 
\affil{$^{1}$Center for Computational Astrophysics, Flatiron Institute, 162 Fifth Ave, New York, NY 10010, USA}
\affil{$^{2}$Department of Physics and Astronomy, Stony Brook University, NY 11794 USA}
\affil{$^{3}$Earth and Planets Laboratory, Carnegie Institution for Science, 5241 Broad Branch Road NW, Washington, DC 20015, USA}
\affil{$^{4}$Univ. Grenoble Alpes, CNRS, IPAG, F-38000 Grenoble, France}
\affil{$^{5}$University of Michigan, 323 West Hall, 1085 S University Ave, Ann Arbor, MI 48109, USA}
\affil{$^{6}$Universidad de Chile, Camino el Observatorio 1515, Santiago, Chile}
\affil{$^{7}$Department of Astronomy, 501 Campbell Hall, University of California, Berkeley, CA 94720-3411, USA}
\affil{$^{8}$European Southern Observatory, Karl-Schwarzschild-Str. 2, 85748 Garching, Germany}
\affil{$^{9}$Institute for Advanced Study, 1 Einstein Drive, Princeton, NJ 08540, USA}
\affil{$^{10}$School of Physics \& Astronomy, University of Leicester, Leicester, LE1 7RH, UK}
\affil{$^{11}$Department of Physics and Astronomy, The University of Georgia, Athens, GA 30602, USA}
\affil{$^{12}$Center for Simulational Physics, The University of Georgia, Athens, GA 30602, USA}
\affil{$^{13}$School of Physics \& Astronomy, University of Leeds, Leeds LS2 9JT, UK}
\affil{$^{14}$Max Planck Institute for Astronomy, K\"{o}nigstuhl 17, 69117, Heidelberg, Germany}
\affil{$^{15}$Center for Astrophysics \textbar{} Harvard \& Smithsonian, 60 Garden Street, Cambridge, MA 02138, USA}
\affil{$^{16}$National Radio Astrophysical Observatory, 520 Edgemont Road, Charlottesville, VA 22903, USA}
\affil{$^{17}$New Mexico State University, Department of Astronomy PO Box 30001, MSC 4500 Las Cruces, NM 88001, USA}
\affil{$^{18}$Departamento de F\'isica, Universidad de Santiago de Chile, Av. Ecuador 3493, Estaci\'on Central, Santiago, Chile}
\affil{$^{19}$School of Physics \& Astronomy, Monash University, Vic. 3800, Australia}
\affil{$^{20}$Leiden Observatory, Leiden University, P.O. Box 9513, NL-2300 RA Leiden, the Netherlands}
\affil{$^{21}$Center for Theoretical Astrophysics and Cosmology, Institute for Computational Science, University of Z\"{u}rich,\\\quad Winterthurerstrasse 190, CH-8057 Z\"{u}rich, Switzerland}
\affil{$^{22}$Lunar and Planetary Laboratory, University of Arizona, 1629 E. University Boulevard, Tucson, AZ 85721, USA}
\affil{$^{23}$Department of Physics and Astronomy, Iowa State University, Ames, IA 50010, USA}
\affil{$^{25}$University of Wisconsin-Madison, 2535 Sterling Hall, 475 N. Charter Street, Madison, WI 53706, USA}
}

\begin{document}
\begin{frontmatter}

\maketitle

\begin{abstract}
A stunning range of substructures in the dust of protoplanetary disks is routinely observed across a range of wavelengths. These gaps, rings and spirals are highly indicative of a population of unseen planets, hinting at the possibility of current observational facilities being able to capture planet-formation in action. Over the last decade, our understanding of the influence of a young planet on the dynamical structure of its parental disk has progressed significantly, revealing a host of potentially observable features which would betray the presence of a deeply embedded planet. In concert, recent observations have shown that subtle perturbations in the kinematic structure of protoplanetary disks are found in multiple sources, potentially the characteristic disturbances associated with embedded planets. In this work, we review the theoretical background of planet-disk interactions, focusing on the kinematical features, and the current methodologies used to observe these interactions in spatially and spectrally resolved observations. We discuss the potential pit falls of such kinematical detections of planets, providing best-practices for imaging and analysing interferometric data, along with a set of criteria to use as a benchmark for any claimed detection of embedded planets. We finish with a discussion on the current state of simulations in regard to planet-disk interactions, highlighting areas of particular interest and future directions which will provide the most significant impact in our search for embedded planets. This work is the culmination of the \emph{`Visualizing the Kinematics of Planet Formation'} workshop, held in October 2019 at the Center for Computational Astrophysics at the Flatiron Institute in New York City.
\end{abstract}

\begin{keywords}
protoplanetary disks -- planet formation -- exoplanet formation -- exoplanet detection methods
\end{keywords}
\end{frontmatter}

\section{Introduction}
\label{sec:introduction}

Since the detection of the first exoplanet, 51~Pegasi~b, in 1995 \citep{Mayor_Queloz_1995}, a stunning variety of planets and planetary system architectures have been detected. With over 4,000 confirmed exoplanets and a comparable number of `objects of interest' waiting to be followed up, we have arrived at the striking conclusion that the Solar System is distinctly atypical \citep[see Fig.~\ref{fig:exoplanet_pop};][]{winn_fabrycky_2015}. It is no longer the case that small, terrestrial planets inhabit the inner regions of planetary systems while the larger gas-giants slowly trundle through the outer regions.

To understand this diversity of planets and planetary systems, we must understand both the planet formation process and the formation environment, the protoplanetary disk. Recent developments in both sub-mm interferometry (e.g the commissioning of the Atacama Large (sub-)Millimeter Array, ALMA) and several instruments pioneering extreme adaptive optics (for example GPI on Gemini, SPHERE on VLT, HiCIAO on Subaru and MagAO(x) on Magellan) have revealed a comparable level of diversity in the physical structure of these potentially planet-hosting disks.

The most interesting possibility opened by these observations is that these disks might already contain young, nearly fully formed planets, giving us a new observational window into exoplanets. Of particular note is the ubiquity of concentric gaps and rings observed in the sub-mm continuum \citep{ALMA_ea_2015, Andrews_ea_2016, Andrews_ea_2018, Long_ea_2018}, in addition to numerous spiral features, primarily detected in the sub-$\mu$m grains in the disk atmosphere \citep{Hashimoto2011, Garufi2013, Benisty2015}. These structures are routinely interpreted as evidence of unseen perturbers lurking within the disks, hinting at the possibility of tracing a population of recently formed planets.

\subsection{Detecting young planets}

\begin{figure*}
\centering
\includegraphics[width=\textwidth]{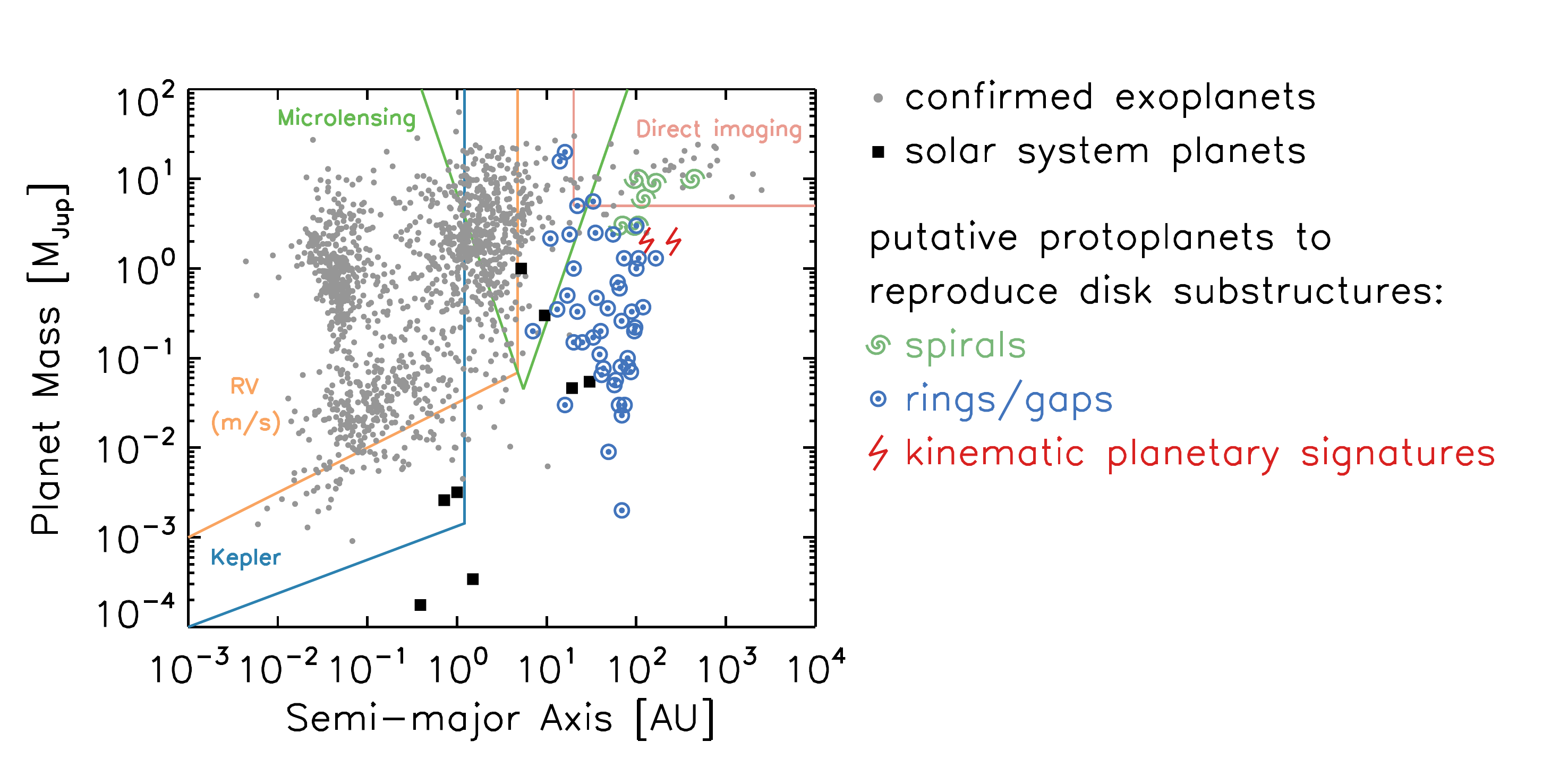}
\caption{The population of exoplanets detected through conventional techniques (grey dots) in comparison with putative young planets in disks (see legend for details). We have also marked on the plot the approximate parameters space accessible to the different techniques (coloured lines). The topic of this review concerns the detection using kinematical information. Updated from \citet{Bae_ea_2018}.}
\label{fig:exoplanet_pop}
\end{figure*}

The most direct way to detect these planets within their natal disk is through the detection of visible and near infrared (NIR) emission, either thermal emission from the gradually cooling planet themselves, or recombination line emission, such as H$\alpha$, indicative of localised accretion onto the planet and / or material falling onto the circumplanetary disk \citep[CPD;][]{Aoyama2018, Thanathibodee_ea_2019, Szulagyi2020}. Extensive campaigns hunting for H$\alpha$ emission \citep[for example,][]{Cugno2019, Zurlo2020}, have only found one system with accreting planets: PDS~70 \citep[PDS~70~b, PDS~70~c,][]{Keppler_ea_2018, Mueller_ea_2018, Wagner2018, Haffert_ea_2019}. However, this source is somewhat special in that PDS~70~b is observed within a significant gap almost entirely devoid of material \citep{Keppler_ea_2019}. This lack of opacity from small grains plays an important role in how readily the planet can be detected in the infrared and suggests that the detection of planets which are still embedded within a disk with no clear gap can be challenging.

An alternative approach is to search for influences of the unseen planet on the physical structure of the disk itself. A long-standing prediction of embedded planets is their ability to open gaps in the gas distribution, inciting a comparable level of substructure in the dust distribution \citep{Lin1986} \footnote{At sub-mm wavelengths, observations of the dust continuum are typically able to be imaged at higher angular resolutions than those of molecular lines as they can exploit the large bandwidth available while line emission will be limited by the intrinsic width of the emission line. As such, much of the `substructure' observed in disks is detected in the continuum.}. Generic simulations of disk-planet interactions can now reliably reproduce observed dust distributions \citep{Zhang2018, lodato_2019}. Figure~\ref{fig:exoplanet_pop} compares the planets used to model the observed substructures in the sub-mm continuum (green and blue symbols) to the confirmed exoplanet population (gray dots). Although concentric rings and gaps are compatible with an embedded planet scenario, they are not unique to that scenario. Various (magneto-)~hydrodynamical instabilities, such as the magneto-rotational instability \citep{flock_2015, bethune2017, riols2019, Riols_ea_2020}, zonal flows \citep{uribe2011}, compositional baroclinic instabilities \citep{Klahr_Bodenheimer_2003} or radially variable magnetic disk winds \citep{Suriano_ea_2018}, have all been shown to produce comparable structure in both the gas and dust. Additionally, direct features of planet-disk interactions are therefore essential in confirming the presence of a planet.

The gaseous component may hold the multidimensional dynamic information required to test competing theories and mechanisms. Although substructures in molecular emission, suggestive of surface density perturbations, have been reported \citep[for example,][]{Isella_ea_2016, Teague_ea_2017, Huang_ea_2018}, attributing these features to changes in the disk physical structure rather than from chemical effects \citep[such as][]{Oberg_ea_2015, Bergin_ea_2016, Cazzoletti_ea_2018} remains a challenge. Even with a suite of molecular line observations, it is hard to infer the physical structure of potential gaps in the gas surface density from the line intensity alone \citep{Facchini_ea_2018}.

With current instrumentation enabling observations at spectral resolutions of $20~{\rm m\,s^{-1}}$, there is an exciting opportunity to search for \emph{kinematical} features, bypassing the complexities of molecular excitation and underlying chemical structures. This possibility was first envisaged by theoretical work that predicted the observable features associated with embedded planets, namely how the locally perturbed velocity fields manifest in molecular line observations \citep{Perez2015, Perez_ea_2018}. Excitingly, the possibility has become reality: two detections of embedded planets have already been claimed, one in HD~163296 \citep{Pinte_ea_2018b} and one in HD~97048 \citep{Pinte_ea_2019}, where localised deviations in the gas velocities are best described by an embedded planet. These are included in Fig.~\ref{fig:exoplanet_pop} as the red lightning bolts. In addition, similar kinematical features have been reported in HD~100546 \citep{Casassus_Perez_2019, Perez_ea_2020} and 8 circumstellar disks observed by the DSHARP program \citep{Pinte_ea_2020}, although dedicated numerical modelling of these additional disks to infer planet masses has not yet been carried out and these objects are not shown in Fig.~\ref{fig:exoplanet_pop}.

\subsection{Constraining the Planet Formation Process}

Detecting planets embedded in their parental disk offers an entirely unique opportunity to study the planet formation process; knowing where the planets are allows for a much more direct relation between properties of the disk and the planet. We give a few more specific examples below.

\subsubsection{Formation Mechanism}

Currently, two modes of planet formation are thought to be viable: a slow process of `core accretion' \citep{Safronov1972}, wherein planets grow from the bottom up, or the rapid collapse of a region of the disk via gravitational instability (GI), directly forming a planet \citep{Boss_1997, Kratter2016}. The former has been shown to account for most of the observed exoplanet population at short (within a few au) separations \citep[e.g.][]{Benz2014}, which is the population accessible through techniques such as radial velocity and transits. However, these mechanisms are exceptionally inefficient at radii of tens of au where dynamical timescales are prohibitively large for the core accretion scenario.

In the last few years, the theory of core accretion has seen a significant revision thanks to `pebble accretion' \citep{OrmelKlahr2010, LambrechtsJohansen2012}. The growth of initial planetary seeds is much faster when considering that disks contain large amount of mm-sized grains (pebbles) rather than considering only km-sized bodies (planetesimals) as in traditional core accretion. This distinction is important as pebbles are subject to gas drag due, while planetesimals only feel gravity and therefore the gravitational focusing will greatly enhance the cross section of the planetesimal and thus enhancing the accretion rate onto it. The caveat is that some other process is needed to jump start pebble accretion by forming initial planetary seeds, required to have a mass comparable to Ceres ($\sim 5 \times 10^{-7}~{\rm M}_{\rm Jup}$). Thanks to these revised timescales, pebble accretion remains a viable mechanism to form planets at large separations on Myr timescale \citep{Johansen2017}.

These outer regions are also those where GI is typically active, in disks where the disk mass, a poorly constrained property at best, must be large enough such that the gravitational potential of the disk is comparable to that of the host star \citep{Clarke2009, Rafikov2009}. However, GI \textit{typically} forms objects much more massive than the majority of known exoplanets. While analytical constraints show that the minimum planet mass that can be formed through GI can be as low as around $1.1~{\rm M}_{\rm Jup}$ \citep{cadman2020} in all parameter space the typical value is more like $\sim 10~{\rm M}_{\rm Jup}$ \citep{forgan2011}. This is in agreement with both parameterised population synthesis models \citep{forgan2013, forgan2018} and high resolution global hydrodynamics simulations \citep{hall2017}. It also agrees with observations. A sample of 199 FGK stars found that between 1\% and 8.6\% of systems, at a confidence level of 95\%, had objects that most likely formed through gravitational instability \citep{vigan2017, Vigan_ea_2020}. This suggests that GI as a planet formation pathway is rare, but it is still statistically consistent with being the dominant formation pathway for rare systems such as HR 8799 \citep{marois2008}, which have massive objects on wide orbits.

Identifying planets forming within $\lesssim 10$~au will therefore provide tight constraints on what dynamical processes must be involved in their formation. Furthermore, the planetary mass can be inferred by matching spectrally resolved kinematic features with hydrodynamical predictions. This {\em dynamical} measurement allows for a highly complimentary approach to masses derived by modelling the observed emission from the embedded planet. With more robust constraints on the mass of embedded planets to hand, a direct test between formation theories is possible; for example, GI tends to form planets with typical masses well above Jupiter, whereas pebble accretion has a preference for Neptune mass planets.

\subsubsection{Planet Composition}

The formation location of the planet determines the chemical composition of the material accreted onto the planet, resulting in highly variable atomic abundances \citep{Oberg2011, Oberg_ea_2016}. Of particular importance, primarily as it can be measured in the atmospheres of exoplanets \citep[e.g.][]{Madhusudhan2019}, is the ratio between the carbon and oxygen atomic abundances, the C/O ratio. With substantial progress made on measuring the local C/O ratio within the disk \citep{Bergin_ea_2016, Cleeves2018, Miotello2019, LeGal2019}, it is in principle possible to trace the C/O ratio in the immediate vicinity of the forming planet.

The planet formation process is a highly complex process. It is still not clear specifically what and from where in the disk is material accreted onto the planet \citep{Booth2017, Ilee2017, cridland19}. Knowledge of the location of the forming planet is therefore essential if spatial changes in the C/O ratio are to be associated with compositional changes in the accreted material. For example, large scale radial variations in the CO abundance, a proxy of the local C/O ratio, are observed in a handful of sources \citep{Zhang2019} in addition to theoretical models suggesting temporal changes \citep[e.g.][]{Krijt2018}. Simultaneously, by characterising the gas velocity structure around the planet, it is possible to provide observational constraints on how the material is being transported from the disk to the planet which can be fed back into simulations of atmospheric formation \citep[e.g][]{Tanigawa2012, Szulagyi2014, Morbidelli_ea_2014}.

\subsubsection{Migration}

For planets embedded within a gas-rich disk, migration is a natural outcome necessary to preserve angular momentum \citep{KleyNelson2012, Baruteau2014}. This process plays an important part in rearranging exoplanetary systems during their formation, which can result in significantly different orbital configurations or system architectures which are inferred through planet-hunting campaigns. The detection of an embedded planet will therefore provide essential evidence with which to better understand the migratory patterns of young planets.

The initial detection of an embedded planet will allow for a better understanding of the formation location assuming that significant migration has not already taken place (see Fig.~\ref{fig:exoplanet_pop}), delivering much-needed initial conditions for models of planet migration. In parallel, the radial dust morphology has been shown to be strongly influenced by migrating planets \citep{Perez_ea_2019a, Meru2019, Nazari2019} and combining these two pieces of information we will be able to characterise the significance of migration.

Understanding the migratory path of a planet is also important to better understand which reservoir materials are accreted and thus how the atmospheric abundances can change \citep{Cridland_ea_2016, Cridland_ea_2017}, linking to the previous point about planet composition. For example, for the case of the Solar System, core formation outside the N$_2$ snowline at $\sim 45$~au and subsequent inward migration has been invoked to account for the abundances of atomic species in Jupiter \citep{Owen_ea_1999, Oberg2019, Bosman_ea_2019}.

\vspace*{0.4cm}  

It is therefore clear that robust detections of embedded, still-forming planets will provide the planet-formation and exoplanet communities with much needed observational evidence for their modelling, and to provide essential context for the planetary demographics which are continually being revised.

\begin{figure*}
    \centering
    \includegraphics[width=\textwidth]{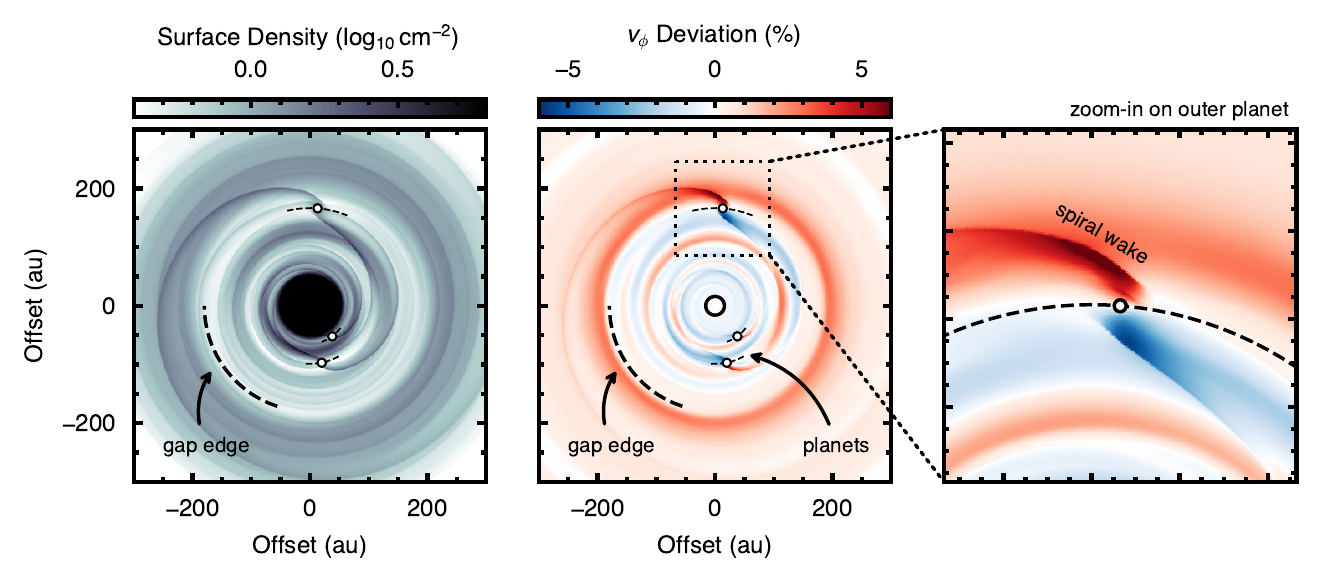}
    \caption{Illustrating how a planet will influence the gas structure and the velocity structure with a 3D hydrodynamical simulation. The left panel shows the gas surface density with three gaps opened up by still-forming planets. The central panel shows the resulting velocity perturbations in $v_{\phi}$, relative to the background rotation velocity. The zoom-in of the outer panel, shown in the right panel, highlights the spiral wakes.}
    \label{fig:planetdiskinteractions}
\end{figure*}

\subsection{Layout of This Article}

This article is the culmination of a recent workshop on \emph{`Visualizing the Kinematics of Planet Formation'}, hosted at the Center for Computational Astrophysics at the Flatiron Institute in New York during October 2019. We aim to present the current state of the field in terms of detecting planets through their influence on the dynamical structure of their parental disk. In \S\ref{sec:methods} we will review the current techniques used to extract velocity information at the meters per second level required to detect planets, highlighting recent results using these methods and detailing current hurdles in the interpretation of these signals. We provide a set of criteria which we believe can be used to claim a detection of an embedded planet or not. \S\ref{sec:imaging} focuses on the problem of imaging high-angular resolution interferometric data, discussing known issues with current methods, and describing alternative approach which are used in neighbouring fields. Much of our intuition is built upon theoretical models and simulations, \S\ref{sec:theory} reviews the current state of the art of the simulations used to interpret the velocity disturbances and what improvements are needed to keep up with the significant leaps forward in quality of observational results. A summary will conclude in \S\ref{sec:conclusions}.

\section{The Kinematical Detection of Embedded Planets}
\label{sec:methods}

Until recently, the search for forming planets has centred on the search for emission associated with the young planet: forbidden transitions indicative of accretion or thermal emission in NIR or sub-mm emission associated with CPDs. Despite some high-profile exceptions, e.g. the PDS~70 system \citep{Keppler_ea_2018, Haffert_ea_2019, Isella_ea_2019}, these methods have proven to be exceptionally challenging, primarily owing to the column of gas and dust from the disk in which the planet is embedded or the small physical sizes expected for the continuum emission associated with the CPDs. With the advent of ALMA, and the possibility to image molecular line emission from the parental disk at both high spatial and spectral resolutions, we can approach the hunt for planets from another angle: by searching for the influence of the embedded planet on the host disk.

To trace the background gas structure, we must move to observations of molecular emission. Despite requiring significantly more sensitive observations than continuum emission (as continuum observations can be made over bandwidths many orders of magnitude broader than line emission), molecular emission enables an entirely different view of the protoplanetary disk. Primarily in that we can probe much larger regions of the disk, radially and vertically, than with dust continuum, but also in that we can trace the \emph{dynamical} structure of the gas. Owing to the significant chemical stratification within the disk (different chemical species require different temperatures, densities and ionisation levels to form and remain abundant enough to be detected), a careful choice of molecular lines allows one to trace distinct vertical regions within the disk and thus trace a full $z/r$ range. A similar approach can be used by selecting a range of less abundant isotopologues such that the lines span a range of optical depths \citep[a common selection are the low $J$ transitions of $^{12}$CO, $^{13}$CO and C$^{18}$O which typically trace $z/r \sim 0.3$, 0.2 and 0.1, respectively; e.g.][]{Pinte_ea_2018a}.

The dynamical structure can be inferred from the Doppler shift of the emission lines. With ALMA able to achieve a 30~kHz resolution, equivalent to $\approx 20~{\rm m\,s}^{-1}$ at 340~GHz ($R \sim 10^7$), the intrinsic line profiles, with typical FWHM $\sim 300~{\rm m\,s}^{-1}$ dominated by thermal broadening, are readily resolved. Offsets between the measured line center and the rest-frame frequencies of these lines can therefore be interpreted as the projected line of sight velocity. By being able to trace the dynamical structure of the gas at a range of heights in the disk, we are able to directly search for the dynamical signatures of the mechanisms which are driving the substructure observed in the dust and hence distinguish between potential scenarios.

\subsection{Theoretical Expectations for Embedded Planets}
\label{sec:methods:theortical_expectations}

In this section, we provide a brief overview of the dominant features expected.

\subsubsection{Gaps}

Embedded planets will excite waves at Lindblad resonances, transporting angular momentum which is deposited into the disk when the waves shock. These shocks exert torques on the local gas, carving a gap, as shown in the left panel of Fig.~\ref{fig:planetdiskinteractions}. The opening of the gaps will modulate the rotation of the gas, $v_{\phi}$, due to the sharp radial gradients in the gas pressure, leading to a hastening of the rotation at the outside of the gap and slowing of the rotation inside \citep{Kanagawa2015, Perez2015, Teague_ea_2018a}, highlighted in the central panel. Grains will drift to local maxima in the gas pressure \citep{Whipple_1972}, opening similar gaps in the dust.

\subsubsection{Spiral Wakes}

Although the planet-opened gaps will result in velocity perturbations around the full $2\pi$ azimuth of the disk, the largest deviations from the background rotation will be along the spiral wakes as shown in the right panel of Fig.~\ref{fig:planetdiskinteractions} \citep{Goldreich1980}. In addition to changes in $v_{\phi}$, the spiral wakes will induce significant radial and vertical velocities \citep{Pinte_ea_2019} as gas is driven away from the embedded planet. These can reach deviations of up to $\pm 10\%$ \citep{Perez_ea_2018, Pinte_ea_2018b, Teague_ea_2018a, Zhang2018, Yun_ea_2019}. Peak deviations depend on background disk properties, such as the disk viscosity or equation-of-state (EOS), as well as planetary mass. Although contemporary studies suggest a clear link between the magnitude of the velocity deviation and the mass of the embedded planet, current simulations lack a full thermodynamic treatment of the shock heating associated with the spiral which may influence this relationship.

\subsubsection{Vertical Motions}

Unlike the large dust grains, those which emit at sub-mm wavelengths, and are confined to the disk midplane, the gas component extends over a large vertical extent over which embedded planets are expected to drive significant motions. The most distinctive of these vertical motions are the `meridional flows' \citep{Kley_ea_2001, Tanigawa2012, Szulagyi2014, Morbidelli_ea_2014}, recently detected by \citet{Teague_ea_2019a}. As the planet opens a gap, material viscously spreads from radii outside the gap towards the gap center, before falling towards the midplane at the gap center in order to maintain hydrostatic equilibrium. This results in a very characteristic flow around an opened gap, with eddies in the gas bounding the gap. It is yet unclear how azimuthally extended these flows are around the forming planet, however most simulations currently suggest that the vertical motions are strongest at the location of the planet. As the velocity of the vertical motions are directly related to the depth of the gap, the detection of gas flows towards the midplane at the radius of the planet would provide strong evidence that there is a significant perturbation in the gas surface density.

Furthermore, \citet{Zhu2012, Zhu2015} demonstrated that when a vertical temperature gradient is present, an embedded planet will excite buoyancy resonances in addition to the Lindblad resonances discussed above. Such resonances will result in lower amplitude over-densities than a Lindblad resonance, but will drive higher velocity perturbations, predominantly in the vertical direction. These additional spirals will have a strong vertical dependence on both their pitch angles (how tightly wound the spirals are). Additionally, the strength of the velocity perturbations which are stronger at higher altitudes. \citet{Teague_ea_2019b} recently observed vertical velocity perturbations in the disk of TW~Hya, tracing a tightly wound spiral morphology, unable to be fit by a Lindblad spiral wake. The coincidence of this feature with a large gap in the gas density \citep{vanBoekel_ea_2017, Teague_ea_2017} is highly suggestive of a planet-induced buoyancy spiral.

\subsubsection{Turbulent Motions}

\begin{figure}
    \centering
    \includegraphics[width=\columnwidth]{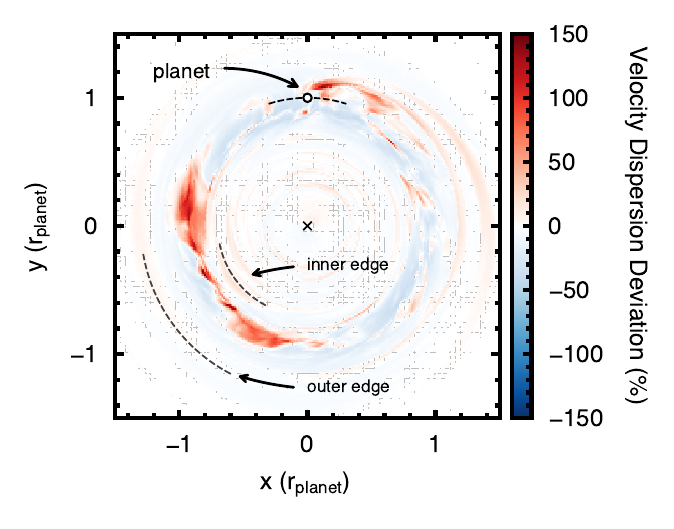}
    \caption{Deviations from the azimuthally averaged vertical velocity dispersion, $\delta v_z$. The marked circle shows the orbit of the planet, rotating in an anti-clockwise direction, while the two concentric dashed lines show the gap edges. The non-thermal velocity dispersions are typically twice the thermal velocity dispersions. Data from \citet{Dong2019}.}
    \label{fig:velocity_dispersion}
\end{figure}

In addition to large, ordered flows, an embedded planet will drive local turbulent motions. In terms of observations, large-scale, ordered flows will result in changes of the line center, while smaller scale, turbulent motions will broaden the observed molecular line. \citet{Dong2019} showed that an embedded $4~M_{\rm Jup}$ planet in a disk with a mass of $0.001~M_{\rm sun}$, induces mildly supersonic vertical velocity dispersions in and around the opened gap, as shown in Fig.~\ref{fig:velocity_dispersion}. While these simulations did show that there was enhanced velocity dispersion along the outer spiral arm, the largest dispersions did not coincide with the location of the planet, but were found around the entire azimuth of the gap. It could therefore be expected that in a planet-opened gap an enhancement in the local line width is observed. However, a significant caveat in these simulations is the assumption of some background viscosity which will scale the amplitude of these dispersions.

\subsubsection{Circumplanetary Disks}

At much smaller scales, material accreting onto the young planet will fall onto a circumplanetary disk. On scales of less than a Hill radius, a circumplanetary disk will form introducing an additional large velocity component to the gas \citep{Kley_1999, Lubow_ea_1999}. Several theoretical works have shown that CPDs can be observed in the dust continuum, however this requires the highest spatial resolution of ALMA in order to resolve the Hill sphere, and extremely deep observations to catch the dust content of these disks \citep{Isella_Turner_2018, Szulagyi2018}. Searches for emission associated to these disks have been made, however currently there are no clear detections \citep[there is a tentative detection a CPD around PDS~70~c, however requires follow-up observations for confirmation;][]{Isella_ea_2019}. However, in the gas \citet{Perez2015} showed that this CPD rotation is sufficiently large to decouple the gas from the background rotation. At lower resolutions, this would manifest as a localised broadening in the emission line, similar the non-thermal broadening described in \citet{Dong2019}, but localised to the location of the planet.

\vspace{0.4cm}  

All these simulations suggest that embedded planets will resulting in numerous observable features which could betray their presence, most notably traced in molecular line emission. In particular, embedded planets should drive substantial and detectable flows in the gas on top of the background Keplerian rotation\footnote{It is important to note that this current set of expected features combines work from several groups with different methodologies. Although the general outlook is consistent, more work is absolutely needed to bring these predictions together in a single cohesive picture}. Such kinematic features potentially provide the most robust tracer as they are decoupled from excitation or abundance-related effects (driven either by the local chemistry for the case of molecular lines, or grain evolution for continuum emission) which could result in false-positive detections of embedded planets. Searching for small deviations relative to this background rotation is far easier than determining the absolute velocity structure of the disk. In the following section, we focus on the kinematical approaches to detecting embedded planets, discussing the current methods of extracting precise velocities structures from observations of molecular line emission necessary to uncovered the signatures of embedded planets.

\subsection{Kinematic Observations of Planet-Hosting Disks: Methods and Techniques}
\label{sec:methods:observations}

\begin{figure*}
    \centering
    \includegraphics[width=\textwidth]{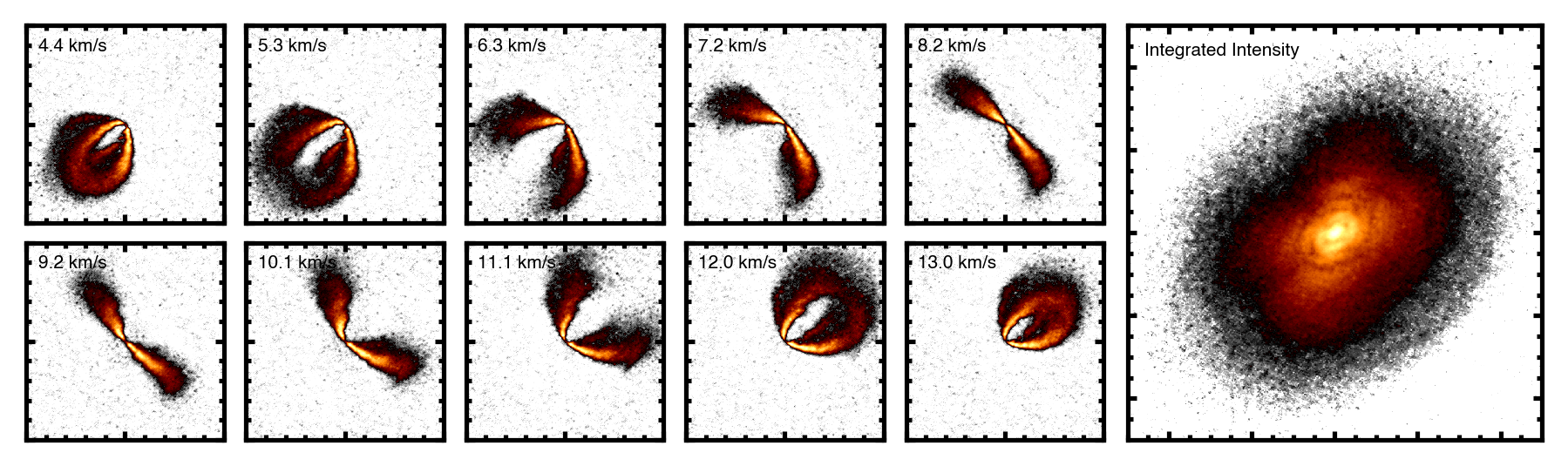}
    \caption{Example channels of $^{12}$CO emission in the disk around HD~163296, with an integrated intensity map (zeroth moment) to the right \citep{Andrews_ea_2018, Isella_ea_2019}. The zeroth moment map was made with a $3\sigma$ clip applied to the data. Note that the ringed structure visible in the zeroth moment map can be attributed to continuum absorption from the far side of the disk and continuum subtraction during the imaging process \citep[e.g.][]{Isella_ea_2018, Keppler_ea_2019}.}
    \label{fig:HD163296_channels}
\end{figure*}

In this section we will focus on line data from a sub-mm interferometer, such as ALMA, although in practice the techniques are applicable to any integral field unit (IFU) like observation (we discuss the additional step and technicalities of imaging the $uv$ data taken with interferometers in \S\ref{sec:imaging}). The end product of an ALMA observation of line emission is a data cube containing three axes (ignoring any polarisation): two spatial, representing the sky coordinates, and the spectral axis. Knowing the rest frequency of the observed line allows for a conversion from frequency to line-of-sight velocity through the Doppler shift of the line. It is more useful to work in velocity-space rather than frequency-space as these values are more directly relatable to the physical processes which we want to understand.

The data cube consists of multiple `channels', each showing the emission integrated over a narrow frequency or velocity range (for typical observations of line emission these ranges are anywhere between 20 and 1000~m\,s$^{-1}$). Figure~\ref{fig:HD163296_channels} shows example channel maps of $^{12}$CO emission in the disk around HD~163296 using the DSHARP data \citep{Andrews_ea_2018, Isella_ea_2019}. The emission is observed in a characteristic `butterfly' pattern owing to the projected rotation of the disk,

\begin{equation}
    v_0(r,\,\phi) = v_{\phi}(r) \sin(i) \cos(\phi) + v_{\rm LSR}
    \label{eq:simple_rotation}
\end{equation}

\noindent where $i$ is the inclination of the disk, $i = 0^{\circ}$ being face-on and $i= 90^{\circ}$ edge-on, $r$ the radius in the disk, $\phi$ being the azimuthal angle measured from the red-shifted major axis of the disk, and $v_{\rm LSR}$ being the systemic velocity of the star. To remove the velocity dependence of the emission, one can integrate along the spectral axis, creating a `zeroth-moment map', as shown in the right-hand, large panel of Fig.~\ref{fig:HD163296_channels}, showing the overall morphology of the line emission.

\begin{figure*}
    \centering
    \includegraphics[width=\textwidth]{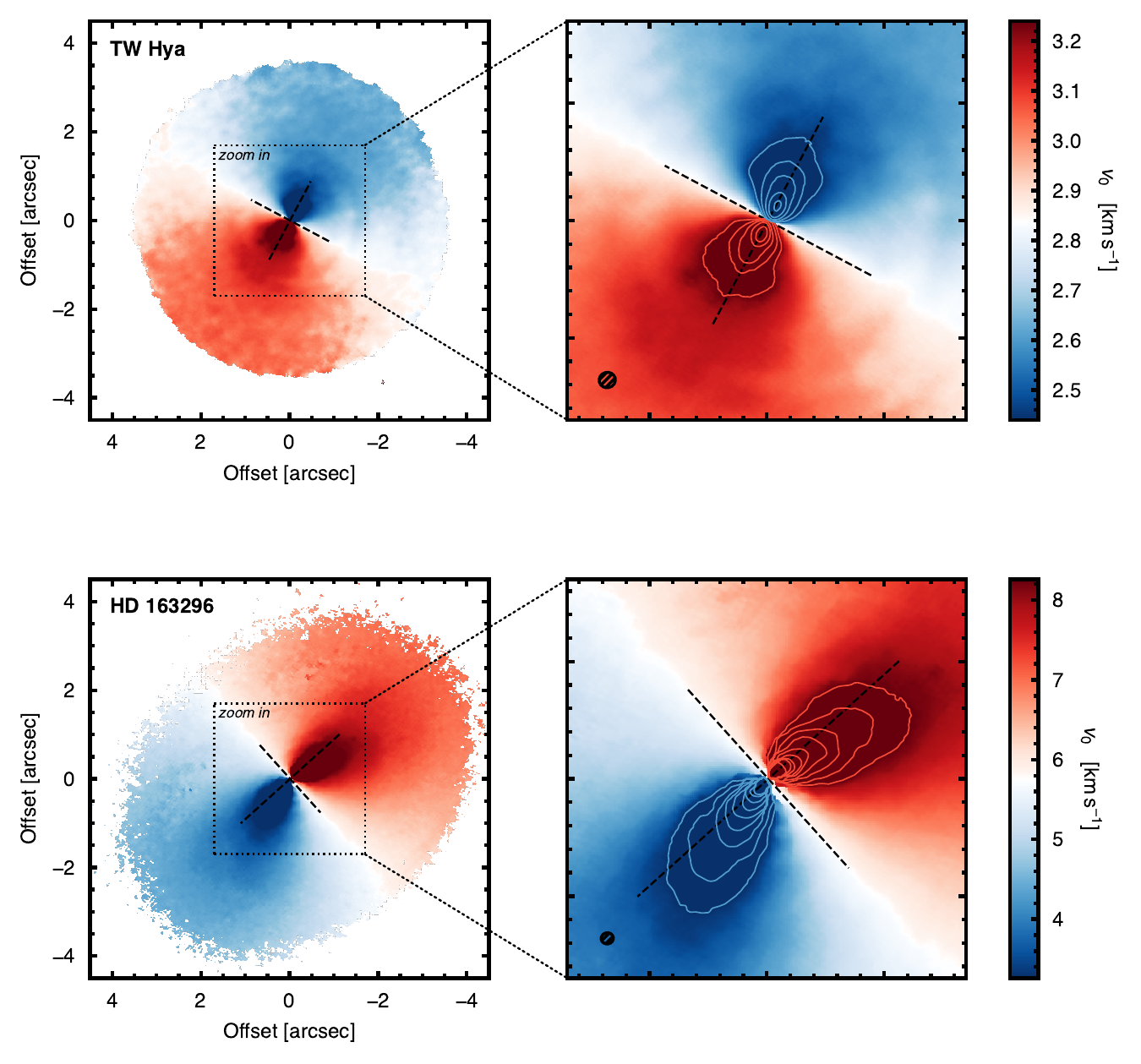}
    \caption{Comparison of the rotation maps for a close to face-on disk, TW~Hya, $i \sim 5^{\circ}$, top, and a moderately inclined disk, HD~163297, $i = 47^{\circ}$, bottom. The maps were created using the `quadratic' method described in \citet{Teague_Foreman-Mackey_2018}. The right panels show a zoom-in of the center of each map with lined contours highlighting the high velocity lobes which bend away from the disk major axis, plotted with dashed lines, due to the elevated emission surface (a `flat' disk would be perfectly symmetric about the major axis of the disk). This effect is far more clearly seen for HD~163296 due to the inclined viewing geometry. The beam sizes are shown in the bottom left of the zoom-in panels as hatched ellipses. The data is taken from \citet{Huang_ea_2018} for TW~Hya and \citet{Isella_ea_2018} for HD~163296.}
    \label{fig:TWHya_rotation}
\end{figure*}

To characterise the background rotation of a disk, the most typical analysis is to make a rotation map, detailing the line measured line center, $v_0$, across the disk as shown in Fig.~\ref{fig:TWHya_rotation}. The dipole morphology arises due the functional form of Equation~\ref{eq:simple_rotation}, with the axis of symmetry being the major axis of the disk. Traditionally these maps are created with an intensity-weighted average velocity, more commonly known as a first-moment map, $M_1 = \sum_i^N I_i v_i \, / \, \sum_i^N I_i$, where $N$ is the number of velocity channels averaged over. This approach, however, is highly sensitive to noise in the data and often requires either user-defined masks or threshold masks which remove `noisy' pixels to produce high quality maps. As such, alternative methods have been advocated for, including fitting a Gaussian component to each pixel \citep[e.g.][]{Casassus_Perez_2019}, fitting a quadratic curve to to the velocity of peak intensity and the two neighbouring velocities \citep[e.g.][]{Teague_Foreman-Mackey_2018}, or just taking the velocity of the peak intensity (called a `ninth-moment' map in CASA, despite bearing no relation to a true statistical moment).

These maps allow one to measure basic geometrical properties of the disk, such as the disk center, inclination, position angle, and, with some assumptions, the dynamical mass of the central star. With higher resolution observations it is also possible to resolve the 3D structure of the disk for the more inclined sources. This manifests as a bending of the dipole lobes away from the semi-major axis (most clearly seen in the HD~163296 panels of Fig.~\ref{fig:TWHya_rotation}). As the lobes appear to bend towards the edge of the disk further from the observer, it is possible to fully determine the orientation of the disk on the sky \citep[see, for example][]{Pietu_ea_2007, deGregorio-Monsalvo_ea_2013, Rosenfeld_ea_2013}.

To measure a radially varying rotation profile, these $v_0$ maps can be azimuthally averaged (most observations of molecular line emission in protoplanetary disks show azimuthally symmetric structure). \citet{Casassus_Perez_2019} used a suite of nested annuli, each taking into account any projection effects of the 3D structure of the disk, to recover a radial profile of $v_{\phi}$ by fitting Eqn~\ref{eq:simple_rotation} to the points in the annuli. \citet{Teague_ea_2018a, Teague_ea_2018b} used a similar method, however rather than fitting the projected $v_0$ value in each pixel, shifted the spectrum in each pixel by the projected disk rotation, $v_{\phi} \sin(i) \cos(\phi)$, such that each spectrum was centred on the systemic velocity, $v_{\rm LSR}$. The aligned spectra could then be compared to find the most appropriate value of $v_{\phi}$ in each annuli. These approaches allow for the inference of $v_{\phi}(r)$ at a precision down to $\approx 10~{\rm m\,s}^{-1}$, depending on the quality of the data. Note that this method has been used to detect weak emission lines in disks but assuming \emph{a priori} the velocity structure, such that lines can be effeciently stacked \citep[e.g.][]{Yen_ea_2016}.

These methods assume that the projected velocities are purely rotational velocities, $v_{\phi}$. However, they are easily extended to additionally account for radial and vertical velocities such that the projected velocity is the superposition of all three projected components,

\begin{equation}
    v_0 = \underbrace{v_{\phi} \sin(i) \cos(\phi)}_{\rm rotational} +
          \underbrace{v_{r} \sin(i) \sin(\phi)}_{\rm radial} +
          \underbrace{v_{z} \cos(i)}_{\rm vertical} +
          v_{\rm LSR}.
    \label{eq:simple_rotation_incl_radial}
\end{equation}

\noindent As demonstrated in \citet{Teague_ea_2019a}, the three velocity components can be disentangled (under the assumption of an azimuthally symmetric velocity distribution) due to their differing dependence on the azimuthal angle, $\phi$. We note that similar methods are used in studies of galaxies \citep[most notably with the \texttt{kinemetry} package;][]{Krajnovic_ea_2006}, which consider a harmonic expansion of circular and radial terms for $v_0$. However, as the intrinsic line widths found in protoplanetary disks are much narrower and systematically broadened due to the finite resolution of the data \citep[e.g.][]{Teague_ea_2016}, these techniques are not directly applicable to protoplanetary disks. \citet{Teague_ea_2019a} was able to use this technique to uncover significant radial and vertical flows, in addition to the previously detected radially varying rotational velocity, in the disk of HD~163296.

The assumption of an azimuthally symmetric disk breaks down when one aims to search for azimuthally localised deviations indicative of an embedded planet. The primary approach for doing this is to search for localised residuals from a background model, $v_0 - \langle v_0 \rangle_{\phi}$ \citep{Perez_ea_2018}. The largest uncertainty here is the choice of background model to use for the subtraction. The most simple approach is to assume a background model that is in Keplerian rotation, optimising the model parameters to find the best fit to the observations \citep[e.g.][]{Walsh_ea_2017, Teague_ea_2019b}. The model can also include considerable amounts of complexity if the data warrants it, such as analytical prescriptions of the emission surface or disk warps, for example, as implemented in the Python package \texttt{eddy} \citep{eddy}. A more complex, or more source-specific, model would be a projection of the azimuthally averaged velocities profiles with a radially varying emission surface as used in \citet{Casassus_Perez_2019}. It is important to consider the flexibility of the model: too flexible and the observations will be over-fit, while if the model is inflexible, large systematic residuals will be produced.

A commonly predicted feature is a `Doppler flip', due to the additional velocity components from the spiral shocks \citep{Perez_ea_2018, Pinte_ea_2018a, Pinte_ea_2018b, Teague_ea_2018a}. This manifests as a positive residual following the outer trailing spiral arm and a negative residual tracing the inner leading spiral arm. Again, due to the projection effects and the superposition of radial and vertical motions also driven by the shock, the strength of this signal can vary significantly as a function of azimuth in the disk \citep[see the appendix of][for example]{Pinte_ea_2019}. Recently, \citet{Casassus_Perez_2019} and \citet{Perez_ea_2020} reported the detection of a significant feature resembling a `Doppler flip' in the disk around HD~100546.

\begin{figure*}
    \centering
    \includegraphics[width=\textwidth]{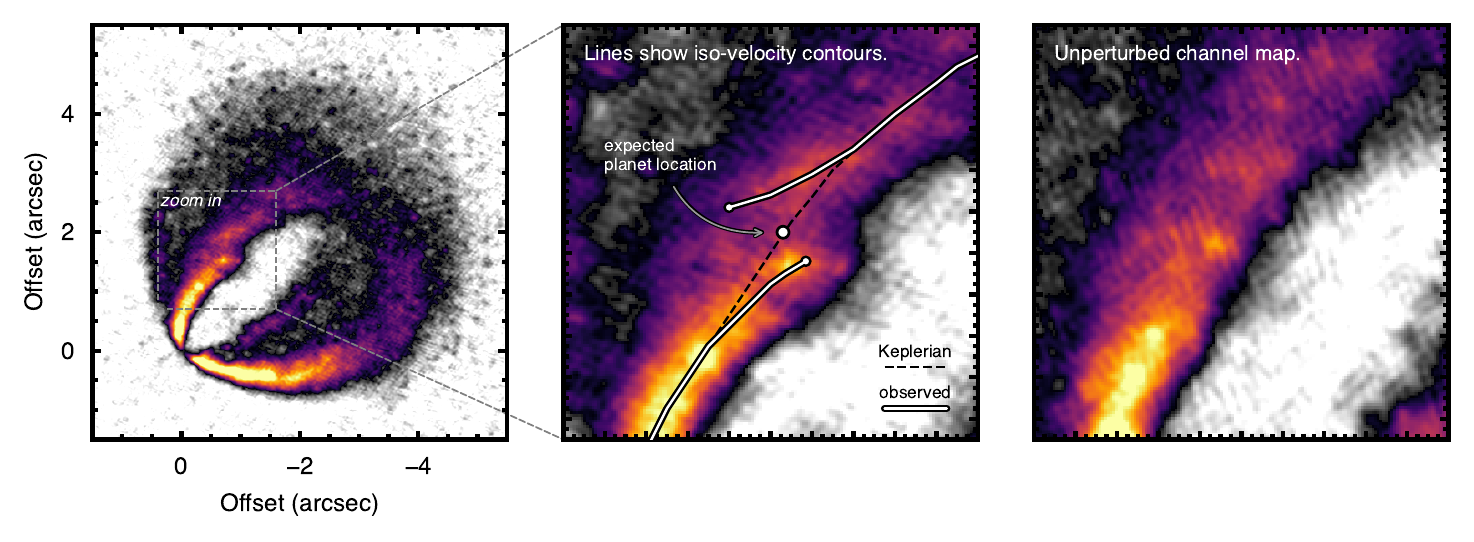}
    \caption{The DSHARP data showing a break in the iso-velocity contours indicative of an embedded planet. The left panel shows the channel at a velocity offset of $1.16~{\rm km\,s}^{-1}$ from the systemic velocity. The central panel shows a zoom-on of the break region. The solid white lines show the observed iso-velocity contour which traces the spine of the emission. The black dotted line shows the iso-velocity contour for Keplerian rotation, revealing a clear departure consistent with a kinematic planetary signature. The right panel shows an unperturbed channel. Note that the predicted planet location in the central panel is the \emph{projected} location at the emission surface of the $^{12}$CO emission.}
    \label{fig:HD163296_planet}
\end{figure*}

Such features are also visible in the channel maps of the data, but manifest in a different way. In a channel map, the emission traces out the iso-velocity contour of the central velocity of the channel. Thus, for a smooth velocity profile the emission is expected to trace out similarly smooth iso-velocity contours. The introduction of velocity perturbations will change the velocity of the gas the emission is tracing and thus shift that emission into an adjacent channel. This manifests as a discontinuity in the iso-velocity contours. Figure~\ref{fig:HD163296_planet} demonstrates this for the feature in HD~163296. Various names have been given to this feature, notably \emph{twist} \citep{Perez2015}, \emph{kinks} \citep{Pinte_ea_2018b, Pinte_ea_2019, Pinte_ea_2020} and \emph{wiggle} \citep{Perez_ea_2018, Perez_ea_2020}. Here we advocate for the umbrella term `kinematic planetary signatures' (KPS) to refer to all these features. Figure~3 from \citet{Perez_ea_2020} provides a comparison of how these feature in the channel maps relate to those observed in the residual maps. \citet{Pinte_ea_2018b} was the first to find such a feature and attribute it to an unseen planetary perturber. Since then, \citet{Pinte_ea_2019} reported a similar feature in HD~97048 and \citet{Pinte_ea_2020} argues for tentative KPSs in nine of the DSHARP sources.

\subsection{Caveats on the Interpretation of Observations}
\label{sec:methods:caveats}

In the previous subsection, we have discussed the various approaches one can take to extract information about the local velocity field and infer the presence of a perturbation relative to some assumed background field. Recent works have shown that these analysis techniques have revealed a stunning array of KPSs, suggesting the presence of unseen planetary perturbers. In this subsection, we discuss the interpretation of these features, in particular their significance and how a strong case can be made for an unseen planet without the option to directly detect it.

One major problem is the inference of an emission height. For optically thick lines, such as the commonly used $^{12}$CO or $^{13}$CO, one expects the $\tau \approx 1$ layer to be relatively narrow such that describing the emission as an elevated 2D surface is appropriate. In this scenario, it is common to deproject the data taking this elevated surface into account, typically using a parameterised description of the emission surface,

\begin{equation}
    z(r) = z_0 \left( \frac{r}{1^{\prime\prime}} \right)^{\psi} - z_1 \left( \frac{r}{1^{\prime\prime}} \right)^{\varphi},
\end{equation}

\noindent as in \citet{Teague_ea_2019b}. Taking $\psi = 1$ and $z_1 = 0$ recovers the conical emission surface described in \citet{Rosenfeld_ea_2013}. However, when the lines become optically thin, as will be the case when aiming to trace closer to the midplane, the $\tau \approx 1$ layer may be extended in the vertical direction, or, for truly optically thin molecules, never reach $\tau \approx 1$ at all. In this regime it may not be applicable to describe the emission layer with such a parameterization, but rather requires the full disk model to be used.

A second hurdle which must be overcome is the interpretation of the residuals in the rotation maps. This is primarily because the projection of the velocities gives rise to azimuthally varying residuals for a constant model offset. For example, an annulus of gas rotating faster than the model predicts will give rise to a residual with an azimuthal modulation of $\cos(\phi)$ \citep{Teague_ea_2019b}. In addition, distinguishing between residuals due to true features in the observations, rather than model misspecification, poses a large challenge. Systematic residuals due to model misspecification will often present as having some order of rotational symmetry. For example, an incorrect inclination will result in a residual with three positive and three negative spokes, while a mis-specified dynamical mass will lead to dipole-shaped residuals. Continuing to explore the parameter space of background models and gain a stronger intuition about systematic residuals that can arise will help in identifying regions of interest.

To compound these issues, the data quality will strongly impact the conclusions that can be drawn from the data. In particular, the long baseline data needed to spatially resolve these features is often noisy over these small spatial scales. This will particularly impact how well an emission surface can be inferred and may be sufficient to shift emission in a channel, yielding a false-positive detection of a localised velocity deviation. A simple solution to this problem is using observations which are designed for kinematic studies: namely very deep, high angular resolution observations with extensive $uv$-plane coverage. However, these are costly in relation to typical continuum observations which can exploit the full bandwidth of ALMA. Alternative techniques for imaging the data, as discussed later in \S\ref{sec:imaging} may additionally help to mitigate these issues.

Ultimately, without a direction detection of a protoplanet, a kinematical detection is not fully confirmed. There is a potential concern that such features are driven instead by hydrodynamical instabilities, such as the vertical shear instability or gravitational instability. Later, in \S\ref{sec:theory}, we discuss our current knowledge of the theory of planet-disk interactions and what simulations need to be developed and run to provide a better set of unique observables.

\subsection{Proposed Criteria for a Kinematical Detection}
\label{sec:methods:criteria}

Given the current techniques, caveats in interpreting images and predictions from simulations, we therefore advocate that the following features are required to claim an embedded planet:

\begin{enumerate}
    \item The detection of a gap, or localised decrease, in the gas surface density. Such a feature would preferably observed in multiple tracers, such as mm-continuum, molecular line emission or scattered light, such that radiative transfer or chemical effects can be ruled out as the cause of the deficit in emission.
    \item The detection of a velocity disturbance which is localised to the centre of the gap. This can be observed in either the rotation map or the channel maps.
    \item The detection of the velocity disturbance in more than one molecular line, tracing different heights within the disk.
    \item The detection in multiple channels of a single line such that the feature is resolved in velocity. This is essential for the modelling required to determine the planet mass.
    \item Enhanced line broadening, ideally coincident with the velocity disturbance, but at least at the same orbital radius.
\end{enumerate}

We note that HD~97048~b \citep{Pinte_ea_2019} has already been ingested into NASA's Exoplanet Database\footnote{\url{https://exoplanets.nasa.gov/exoplanet-catalog/7503/hd-97048-b/}} with an assigned detection method of `disk kinematics' although currently this only satisfies the first two proposed points.

In order to roll out this search for embedded planets to a larger source sample, it is important to understand the scale of observations needed. A typical observation for planet hunting assuming a $1~M_{\rm Jup}$ planet at a separation of 100~au would require $0.1^{\prime\prime}$ spatial resolution (15~au at typical source distance of 150~pc) and a spectral resolution of $100~{\rm m\,s^{-1}}$. For moderately inclined disks, $i \gtrsim 20^{\circ}$, this would be sufficient to detect the localised deviations shown in Fig.~\ref{fig:HD163296_planet} and agree with the proposed criteria for a kinematic detection. As many of these sources are very extended, therefore requiring multiple array configurations to prevent spatial filtering, such observations would require 10 to 20 hours depending on the brightness of the target molecule (note that much of this time is required to well sample the $uv$ space, required for high-fidelity imaging, discussed below, rather than for sensitivity reasons). While this represents a significant investment in terms of observation time, particularly for less abundant species that would trace closer to the midplane where the signatures are stronger, we stress that these are absolutely achievable with standard proposals with ALMA and should be strongly encouraged.

\section{Robust Interpretation of Synthesized Images}
\label{sec:imaging}

In the previous section we have discussed some of observable features which we can be associated with embedded planets. However, these features are subtle and are pushing the limits of what can be discerned with contemporary interferometric data. In particular, as the spatial and spectra scales associated with these features are comparable to the spatial and spectral resolution of the data it is essential to make sure that artefacts are not introduced in the imaging of interferometric data.

An important aspect of working with interferometric data is the consideration of how the images were synthesised. Recent results from the Event Horizon Telescope have widely publicised many of the issues associated with imaging interferometric data and the features that such processes can inject into the final image \citep[e.g.][]{eht-IV-2019}. In this section we provide a short overview of the general problem faced when imaging interferometric data, and propose alternative checks which can be made, such that features observed in channel maps can be attributed to embedded planets rather than imaging artefacts.

\subsection{The Visibility Plane}
\label{subsec:visibility}

Interferometers like ALMA sample the visibility function ${\cal V}$ of a source at a set of discrete spatial frequencies $(u,v)$ fundamentally dictated by the array configuration and observing frequency \citep{thompson_2017}. The visibility function is given by the Fourier transform of the sky brightness distribution $I(l,m)$,
\begin{equation}
    {\cal V}(u,v) = \int \int I(l,m) \exp \left \{- 2 \pi i (ul + vm) \right \} \, \mathrm{d}l\,\mathrm{d}m.
\end{equation}
where $l$ and $m$ are the direction cosines on the sky corresponding to R.A. and declination.
The fundamental interferometric data product is then a set of $N$ calibrated visibility measurements $\boldsymbol{V} = \{V_i\}^N_{i=1}$ at various coordinates in the $uv$-plane (spectral line observations also have an additional frequency ($\nu$) dependence). These visibilities are complex-valued numbers with Gaussian measurement uncertainties proportional to the thermal system noise. Historically, models have been fit in this native Fourier plane: a sky-brightness model $I(l, m\,|\, \boldsymbol{\theta})$ is Fourier transformed to $V(u,v\,|\,\boldsymbol{\theta}$), evaluated at the same spatial frequencies as the observations, and assessed using the (log-)likelihood function of the data
\begin{equation}
    \ln {\cal L} = \ln p(\boldsymbol{V} | \boldsymbol{\theta}) \propto -\frac{\chi^2(\boldsymbol{\theta})}{2}.
    \label{eqn:log-likelihood}
\end{equation}
For unresolved or marginally resolved sources, simple models like elliptical Gaussians are adequate and have the added benefit that $V(u, v\,|\,\boldsymbol{\theta})$ can be specified analytically.

This analysis workflow is also appropriate for more complex astrophysical models, such as channel maps of CO protoplanetary disk emission generated by radiative transfer codes. The benefit of bringing these inherently image-plane models to the visibilities \citep[via the FFT and band-limited interpolation;][]{schwab1984} is that the full information content of the dataset is utilised in assessing the probability of the model parameters $\boldsymbol{\theta}$ within a Bayesian framework and ``nuisance'' parameters can be marginalized out of final parameter estimates. For example, the Keplerian velocity pattern of a protoplanetary disk can be used to precisely constrain the mass of the central star(s), which dominates the gravitational potential, while marginalizing over disk structure parameters \citep[e.g.,][]{czekala2017b}. Directly fitting the measured visibilities has the benefit of avoiding beam sidelobe effects \citep[e.g., the multiple rings of AS~209;][]{guzman2018} and calibration artefacts can often be effectively remedied in a self-consistent manner \citep[e.g., phase self-calibration;][]{hezaveh2013}. The main drawback of visibility-plane fitting, however, is that it can be difficult to assess whether one has specified a sufficient model for the application at hand. Because the Fourier transform carries localised flux in the image plane to many spatial frequencies in the Fourier domain, a good visibility model must necessarily reproduce the observed emission at all positions in the disk and at all observed frequencies. This is challenging when optically thick and optically thin molecular line and continuum emission and absorption are present. For the purposes of detecting the kinematic signature of a protoplanet in a disk, correctly modeling the photodissociation layer, molecular abundance, and vertical temperature gradient may be second-order concerns relative to the (primarily Keplerian) velocity field. Yet, these features must be correctly modeled in order to produce a good-fitting visibility model, since signal from these features will appear at many spatial frequencies.

For several years now, ALMA has been observing protoplanetary disks with sufficient coverage at high spatial frequencies to produce richly detailed images indicative of gaps, rings, and a multitude of disk substructure. By shifting analysis to the image plane, geometrically motivated models of optically thick emitting surfaces and azimuthally-symmetric velocity fields can bypass many of these fine-grained modeling concerns by building localised model complexity where required. Difficult-to-model but otherwise irrelevant regions of the disk can simply be masked from the analysis.

Because interferometers do not sample all necessary spatial frequencies (and those that they do are corrupted by noise), constructing images, here represented as a flattened array of pixels, $\boldsymbol{I} = \{I_i\}_{i=1}^M$, from the visibility samples $\boldsymbol{V}$ requires making assumptions about the unsampled spatial frequencies. Shifting the analysis from the Fourier plane to the image plane also shifts concerns from \emph{model mis-specification} (i.e., is the model parameterization $I(l, m\,|\, \boldsymbol{\theta})$ sufficiently complex to capture the sky-brightness distribution?) to \emph{image fidelity} (i.e., is the image $\boldsymbol{I} = \{I_i\}_{i=1}^M$ a faithful representation of the true sky brightness $I(l,m)$?). Here we focus on examining the potential pitfalls of image-based analysis and present a robust analysis workflow to promote scientifically valid inferences about kinematically induced features in protoplanetary disks.

\begin{figure*}
    \centering
	\includegraphics{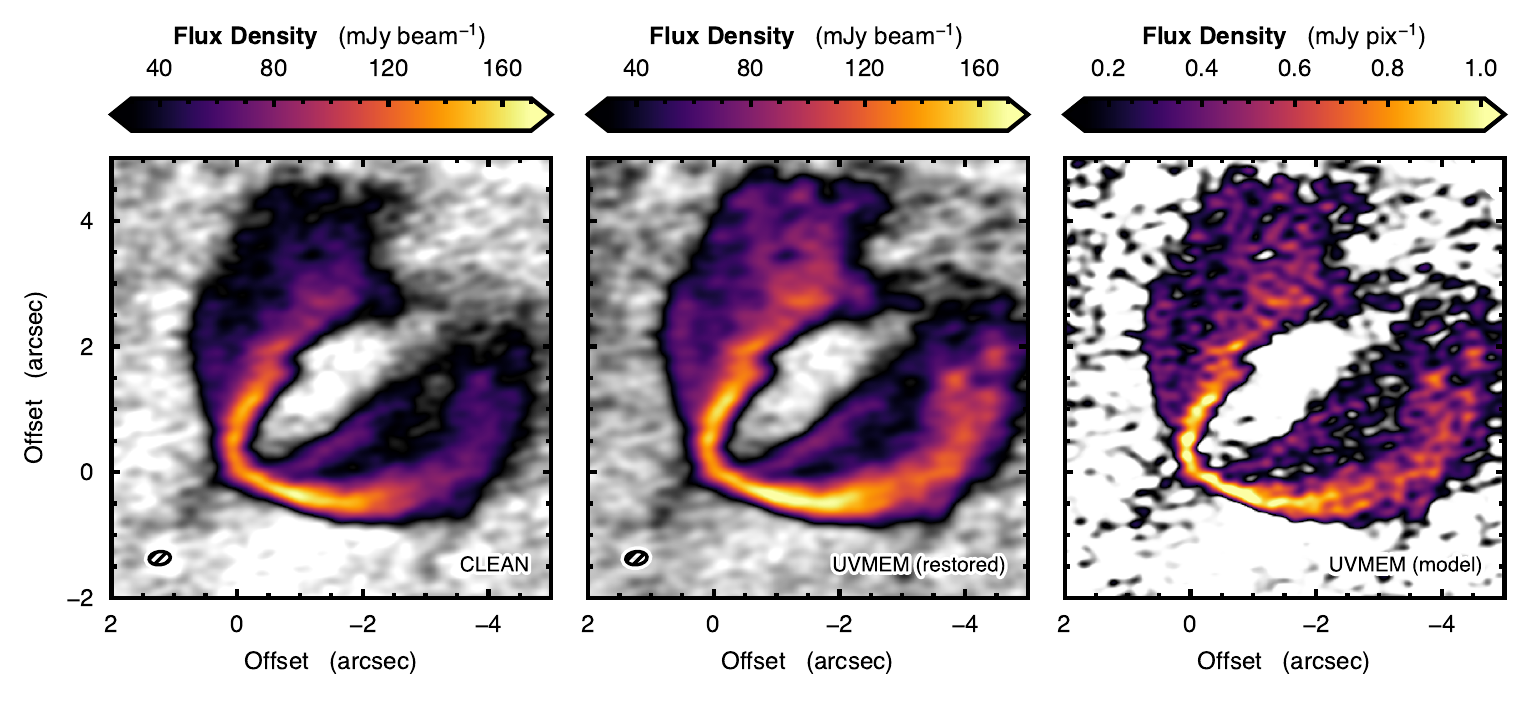}
	\caption{Comparison of a typical \texttt{CLEAN} image, \emph{left}, and a \texttt{uvmem} image, both restored with the same beam as in the \texttt{CLEAN} image, \emph{center}, and the native model image, \emph{right}. The data was originally presented in \citet{Isella_ea_2016} and the \texttt{uvmem} imaging was performed with \texttt{gpuvmem}. While analysis can be performed on the \texttt{uvmem} model directly, any comparisons with the \texttt{CLEAN} images must be performed with the restored image which is the model image convolved with the \texttt{CLEAN} beam.}
	\label{fig:HD163296_comparison}
\end{figure*}

\subsection{CLEAN}
\label{subsec:CLEAN}

The basic spiral configuration of the ALMA array ensures dense $uv$-plane sampling for a majority of array configurations \citep{wootten2009}, even for short exposure times (and thus Earth rotation). In the most extended array configurations (which yield the highest spatial resolution), however, the density of $uv$-plane coverage drops, and care must be taken when synthesising images \citep[Fig.~2 of][neatly demonstrates the relation between $uv$ coverage, the resulting synthesised beam and the clean image]{Andrews_ea_2018}. The more that the observational setup (i.e., the array configuration, observing frequency, and diversity of hour-angle execution) can be optimised to sample the $uv$-plane, the less that algorithmic complexity will be needed to synthesise quality images. Sometimes, due to low source elevation or the expense of integrating longer to achieve Earth rotation, sparse $uv$ coverage is unavoidable.

The CLEAN family of algorithms work by iteratively deconvolving the dirty image (the inverse Fourier transform of the sampled visibilities) with the dirty beam (the inverse Fourier transform of the window function of the array) to build up a model image. In its simplest implementation, the model image is represented by a series of point sources \citep[e.g.,][]{hogbom1974}; more advanced algorithms use multi-scale basis sets \citep[e.g., a series of approximately Gaussian functions of varying widths;][]{cornwell2008}. The deconvolution process is typically implemented as a \emph{procedure}, whereby the regions to deconvolve are identified by the user with a CLEAN mask. The final CLEANed image will often look different depending on the choices of procedural parameters like loop gain and threshold level, underscoring the point that in such a framework there is no one optimal image, but rather a range of many possible images consistent with the data.

The spatially resolved molecular emission from protoplanetary disks, at once diffuse and concentrated (e.g., Figure~\ref{fig:HD163296_comparison}), is generally not well-matched by the basis sets available in CLEAN algorithms today. Adequate images can sometimes be produced by extensive tuning of the procedural deconvolution loop at low gain levels such that enough Gaussians and/or point sources of varying amplitudes are collected in the right proportions. However, this approach is far from optimal when the emission is faint, diffuse, and the $uv$ coverage is sparse. The mismatch between the actual dirty beam and the CLEAN beam (an elliptical Gaussian fit to the core of the dirty beam; used to restore a CLEAN model) makes attaining a decent CLEAN model difficult in practice without substantial user intervention \citep[e.g.][]{Jorsater_vanMoorsel_1995, Walter_ea_2008, Pinte_ea_2020}. This can have disastrous consequences for image fidelity, especially at high dynamic range. Development is needed to identify more suitable CLEAN basis sets and regularization terms.

\subsection{Regularized Maximum Likelihood (RML) Methods}
\label{subsec:RML}

An alternate family of imaging algorithms are the ``maximum entropy'' techniques \citep{cornwell1985,narayan1986}, or, more generally, regularized maximum likelihood methods \citep{eht-IV-2019}. These techniques can be viewed through the same forward modeling framework as in \S\ref{subsec:visibility} with the model being the set of pixel values comprising the image itself
\begin{equation}
    \ln {\cal L} = \ln p(\boldsymbol{V} | \boldsymbol{I}) \propto -\frac{\chi^2(\boldsymbol{I})}{2}.
\end{equation}
Because the Fourier transform is a linear operator, it should be noted that one maximum \emph{likelihood} image is simply the inverse Fourier transform of the visibilities, i.e., the dirty image. Equivalent images that also maximize the likelihood are those that contain emission on spatial frequencies not sampled by the array. In addition, some implementations model the logarithm of the image \citep[e.g.,][]{junklewitz2016}, making the likelihood function itself a non-linear function of the model parameters. The advantage of \emph{regularized} maximum likelihood techniques is that priors $p(\boldsymbol{I})$ may be specified (also called regularization penalty terms) to promote certain image characteristics away from the dirty image. Together, the likelihood and prior yield the image posterior
\begin{equation}
    \ln p(\boldsymbol{I} | \boldsymbol{V}) \propto \ln p(\boldsymbol{V} | \boldsymbol{I}) + \ln p(\boldsymbol{I}).
    \label{eqn:posterior}
\end{equation}
Much of the variety among regularized maximum likelihood methods boils down to choice of the prior, $p(\boldsymbol{I})$. The most common prior in use is the image ``entropy,''
\begin{equation}
    \ln p(\boldsymbol{I}) = - \lambda \sum_i \frac{I_i}{G} \, \ln \frac{I_i}{G} + \mathrm{constant}
\end{equation}
where $G$ is a constant approximately equal to the thermal noise in the image, and $\lambda$ is a scaling factor that controls the relative importance of the likelihood function and the entropy prior \citep{carcamo2018}. This form of prior also enforces strict positivity on the image  pixel values, such that $I_i > 0, \forall\, i$. Image entropy is analogous to the same concept in statistical mechanics, where entropy is defined using the number of possible equivalent microstates of a macroscopic system. In the imaging case, an image with larger entropy is one that would look similar across as many possible pairings of specific pixels $i$ with intensity values $I_i$; a constant intensity image is one with maximal entropy \citep{narayan1986}. Note that the entropy prior alone does not directly enforce a constraint on the image spatial ``smoothness,'' but in practice the entropy prior frequently promotes smoother solutions than CLEAN algorithms because more uniform (and therefore smoother) emission has a higher entropy than point source emission against an otherwise blank background. Of course, priors that do explicitly favor spatial smoothness can also be used in addition to the entropy prior \citep[e.g.,][]{eht-IV-2019}.

The entropy prior makes the full image posterior (Equation~\ref{eqn:posterior}) non-linear, such that the maximum \emph{a posteriori} image must be found via optimization rather than direct inversion. With GPU-accelerated maximum entropy implementations \citep[\texttt{GPUVMEM} \footnote{\url{https://github.com/miguelcarcamov/gpuvmem}};][]{carcamo2018}, it is now computationally tractable to generate large image cubes of protoplanetary disk molecular line emission (Figure~\ref{fig:HD163296_comparison}).

\begin{figure*}
    \centering
    \includegraphics[]{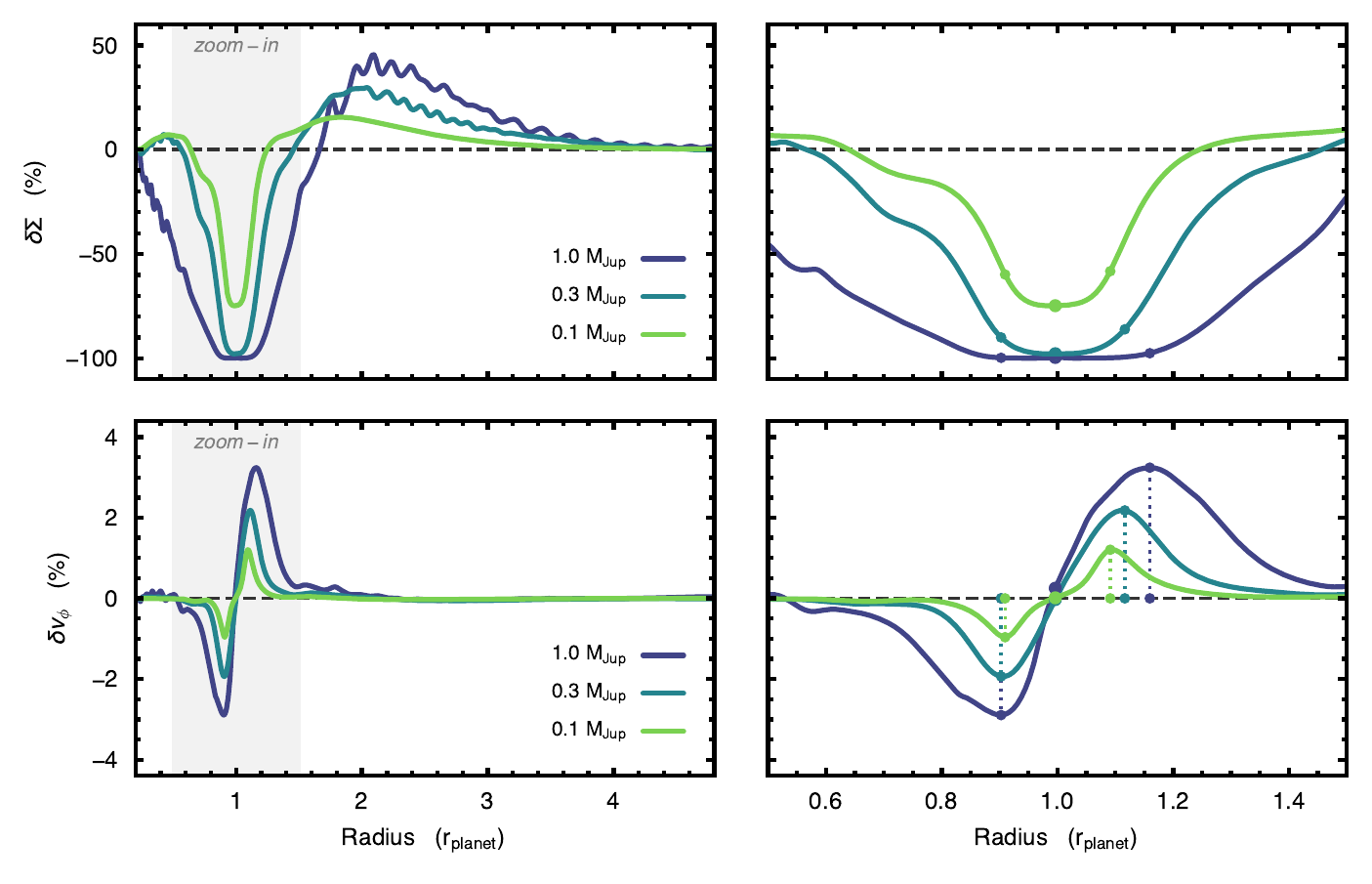}
    \caption{Comparing the shape of the perturbation in the gas surface density, \emph{top}, and the induced deviations in the azimuthal velocity profile, \emph{bottom}, for three different planet masses: 0.1, 0.3 and 1~$M_{\rm Jup}$. The \emph{right} panels show a zoom-in of the \emph{left} panels with the maximum and minimum velocity deviations for each model marked by circles. A more massive planet opens a wider and deeper gap in the surface density, resulting in more significant deviation in velocity spanning a wider range of radii.}
    \label{fig:surface_density_velocity_comparison}
\end{figure*}

\subsection{Recommended Workflow for Assessing Image Fidelity}

With this in mind, we propose that for many of the subtle features which can be associated with an embedded planet, a test of robustness of the feature can be to image the data using an alternative imaging scheme. Although such methods are not implemented in the commonly used \texttt{CASA} package, there is a growing number of community-developed packages to perform such imaging: \texttt{GPUVMEM} \citep{carcamo2018}, \texttt{PRIISM} \citep{Nakazato_ea_2019} and \texttt{MPoL} \citep{mpol}, to name a few. If such packages are unavailable, different imaging schemes, for example different values of Briggs' robust parameter, would provide first-order check that features are real \citep[see the Appendix in][for example]{Pinte_ea_2020}.

Understanding how far one can ``trust'' their synthesised images is vital to reaching valid scientific conclusions. One powerful, though potentially time-consuming, way to assess image fidelity is to construct an imaging cross-validation framework \citep[e.g.,][]{eht-IV-2019}. In such a framework, several mock datasets are created using simulated disk emission structures from which synthesised observations are generated with realistic ALMA array configurations and integration times. A fraction ($\sim 80\%$) of these mock datasets is used to test various imaging strategies and optimise imaging parameters to yield the ``best'' synthesised images, where best is defined relative to the application at hand. For the purposes of detecting protoplanets via their kinematic perturbations, a reasonable figure of merit would be the preservation of any non-Keplerian structures and resilience against introducing any artificial features which could be mistaken for perturbations. Finally, the systematic error budget of the tuned imaging algorithms are assessed using the remaining fraction ($\sim 20\%$) of the mock datasets that were not used in training and propagated into analysis of the disk velocity field. Such a cross-validation methodology can be used in both \texttt{CLEAN} and \texttt{RML} frameworks.

\section{Theoretical Considerations}
\label{sec:theory}

From the plethora of observations made by ALMA (as well as other facilities, e.g., SPHERE on VLT), planet formation within the context of protoplanetary disks is clearly an observationally driven field at the present time. From the theorist viewpoint, the time is ripe to begin to make sense of the complex data acquired from such observations and piece together a cogent picture of planet formation and protoplanetary disk evolution. Until now, most theoretical predictions have revolved around the structures observed in the sub-mm dust continuum and NIR scattered light. In this section, we outline recent results from the theory community and the implications of these results, in particular those relating to the dynamical structure of the disk, for current and future observations of disks.

\subsection{Gap Opening}

Planets are arguably the favoured explanation for the ring and gap structures in protoplanetary disks \citep{Dipierro_ea_2015, Zhang2018}. This is not only because there is now strong evidence that planets are a common occurrence in extra-solar systems \citep[e.g.,][]{Fulton2018, Mulders2018, Hsu2019}, but also because gap-opening by planets is a simple and robust mechanism that is almost unavoidable.

Planets excite waves at Lindblad resonances through gravitational planet-disk interaction \citep{Goldreich1980}, which enable the transport of angular momentum through the disk. Waves with lower amplitudes, i.e. those excited by smaller planets, have to travel further until they steepen and shock, depositing the angular moment back into the disk, but regardless, they all do so eventually \citep{Rafikov2002}. The gas outside the planet's orbit flows outward as these shocks exert a positive torque, while the gas inside the planet's orbit flows inward because the torque is negative. A gap consequently forms centered on the planet’s orbit.

This process can only be countered by diffusion mechanisms that act to fill the gap back in. A typical source of diffusion is turbulence, but the level of turbulence in protoplanetary disks may be sufficiently low, as suggested by some observations \citep{Flaherty2015, Flaherty2017, Pinte2016, Teague_ea_2016, Teague_ea_2018b}, that it should not be able to prevent planets from opening gaps, except for perhaps the smallest, Earth-sized planets.

In the absence of diffusion, technically all planets open gaps, but they clear their gaps at different rates. This rate scales with the planetary torque, which scales with the planet mass squared, and so can become exceedingly long compared to the disk lifetime for the smallest planets. For this reason, it is practical to think of gap-opening planets as the more massive ones, like super-Earths and above, but one should bear in mind that smaller planets are not necessarily excluded from candidacy.

Numerical studies have looked into the properties of planetary \emph{gas} gaps \citep{Lin1986, Bryden_ea_1999, Crida2006}. For the depth of the gap, as a function of the planet-to-star mass ratio $q$, local disk aspect ratio $h/r$, and the viscosity parameter $\alpha$, studies have converged to an answer \citep{Duffell2013, Fung2014, Kanagawa2015, Dong2017a}:

\begin{equation}
    \frac{\Sigma_0}{\Sigma_{\rm gap}} - 1 \approx 0.043\, q^2 \left(\frac{h}{r}\right)^{-5} \alpha^{-1} \, ,
\end{equation}

\noindent where $\Sigma_0$ is the surface density of the disk, and $\Sigma_{\rm gap}$ is the surface density of the gap. This scaling can be understood as the global balance between the one-sided Lindblad torque and the disk viscous torque \citep{Fung2014}. The width of the gap is more difficult to determine because there is no definite edge to the gap. Depending on the definition, different scalings can be empirically derived from simulations \citep{Kanagawa2015, Dong2017a}. These results are most applicable to isothermal disks, and refer only to the gaseous component.

The properties of planetary gaps in the dust distribution are less understood, despite being probed by continuum observations, primarily owing to the poorly constrained coupling of the dust to the gas. Some efforts have been made to systematically study the depth and width of dust gaps \citep{Paardekooper_Mellema2004, Dipierro_ea_2015, Dipierro_ea_2016, Dipierro_Laibe2017, Dong2017b, Zhang2018}. A key point is to distinguish between gaps carved in both gas and dust, and gaps carved only in dust \citep{Dipierro_ea_2016}. In both cases, however, the results depend not only on dust properties, which are poorly constrained, but also on gas properties, assumed thermodynamics \citep{Miranda_Rafikov_2019, Facchini_ea_2020}, the assumed mechanism for angular momentum transport. The origin of turbulence, if present, remains unclear \citep[see review by][]{Lyra_Umurhan2019}. Few, if any, simulations have been performed of gap opening assuming angular momentum transport via winds, which may change the picture more substantially.

\begin{figure*}
    \centering
    \includegraphics[width=\textwidth]{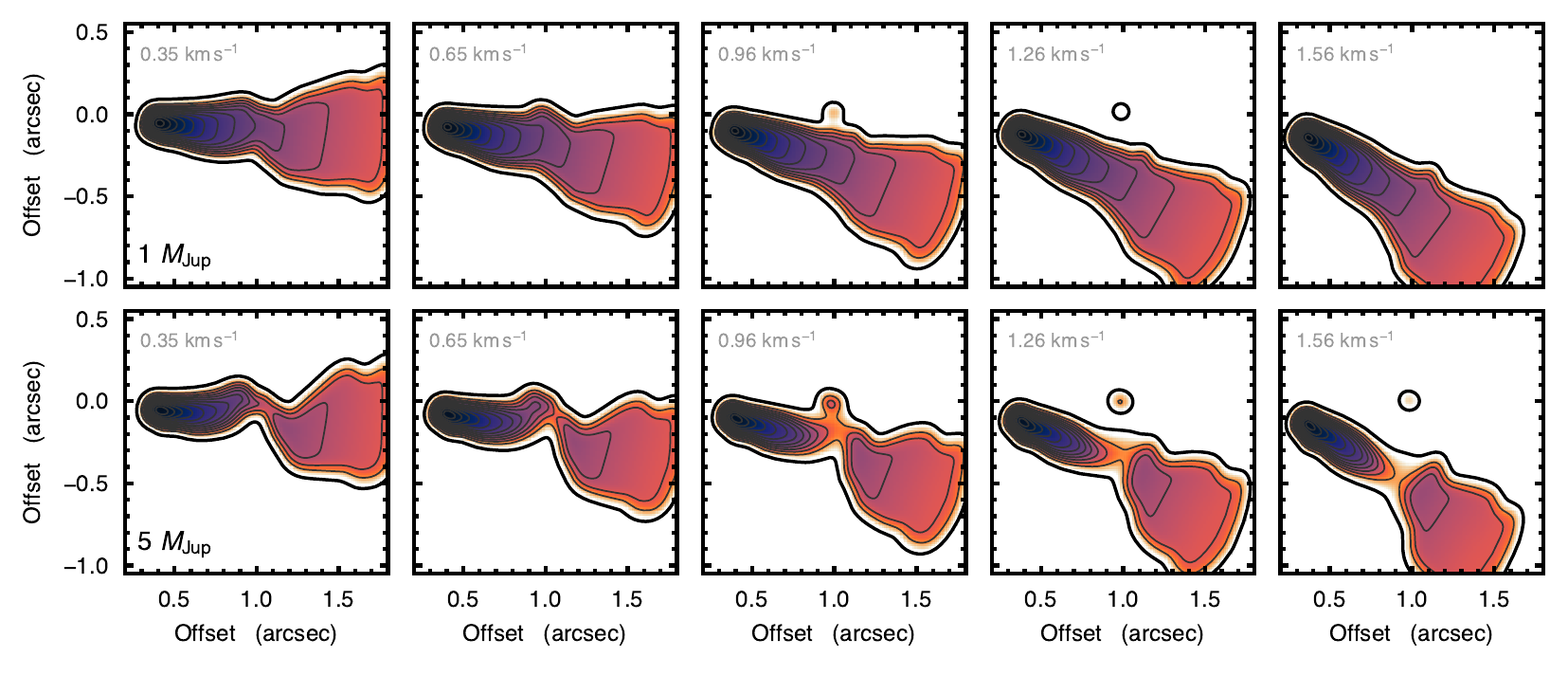}
    \caption{Simulated channel maps of $^{13}$CO emission from disks with a $1~M_{\rm Jup}$ planet, top, and a $5~M_{\rm Jup}$ planet, bottom using the simulations from \citet{Perez2015}. The panels on the left show the characteristic twist in the emission morphology due to the local deviations in velocity due to the spiral wake of the embedded planet while the three right-most panels show an offset peak in emission from the CPD. Clearly the deviations are more significant for the larger $5~M_{\rm Jup}$ planet.}
    \label{fig:KPS_Perez}
\end{figure*}

Following the ideas presented in \citet{dullemond2018b}, \citet{Rosotti_ea_2020} demonstrated that the changes in $v_{\phi}$ observed with current data can be used to quantify the changes in the local gas pressure gradient and thus the change in gas density, as shown in Fig.~\ref{fig:surface_density_velocity_comparison} \citep[see also][]{Teague_ea_2018b}. Comparison with the dust distribution traced by the mm~continuum allows for unique constraints on the coupling of the dust to the gas, enabling a far more robust probe of the depth and width of gap. Furthermore, observations of the molecular line emission with spatial resolutions matching that of the continuum emission will also allow for kinematical probes of the \emph{shape} of such gaps, providing unique constraints on both the gap-opening mechanism or the mass of the embedded planet \citep[e.g.][]{Kanagawa2015, Zhang2018, Yun_ea_2019}.

Another key aspect of planetary gaps in the context of disk kinematics is the idea that there must be meridional circulation, movement in the $r$-$z$ plane, present in them. 3D simulations have revealed that the disk is torqued by the planet more strongly in the midplane than at higher altitudes \citep{Szulagyi2014, Fung_Chiang2016, Pinte_ea_2019}. To reach steady state, disk gas must cycle through different altitudes and obtain a time-averaged torque that is net zero at all heights. This circulation may be already observed \citep{Teague_ea_2019b}. Although turbulent motion is present all around the gap \citep{Szulagyi2014, Fung_Chiang2016, Dong2019, Teague_ea_2019b}, it is fastest near the planet where the torque is strongest; in other words, the meridional circulation should be azimuthally localized. This meridional flow around gap edges is distinct from the planetary accretion flow, which is also meridional; we discuss that meridional flow in Section~\ref{sec:theory:meridional_flows}. Future observations that measure this, in particular those tracing different heights in the disk by targeting molecular line emission with different optical depths, would directly locate the planet, and be the definitive proof of the planetary origin of the gap.

\subsection{Velocity Perturbations from Spirals}
\label{sec:theory:local_perturbations}

\begin{figure*}
    \centering
   \includegraphics{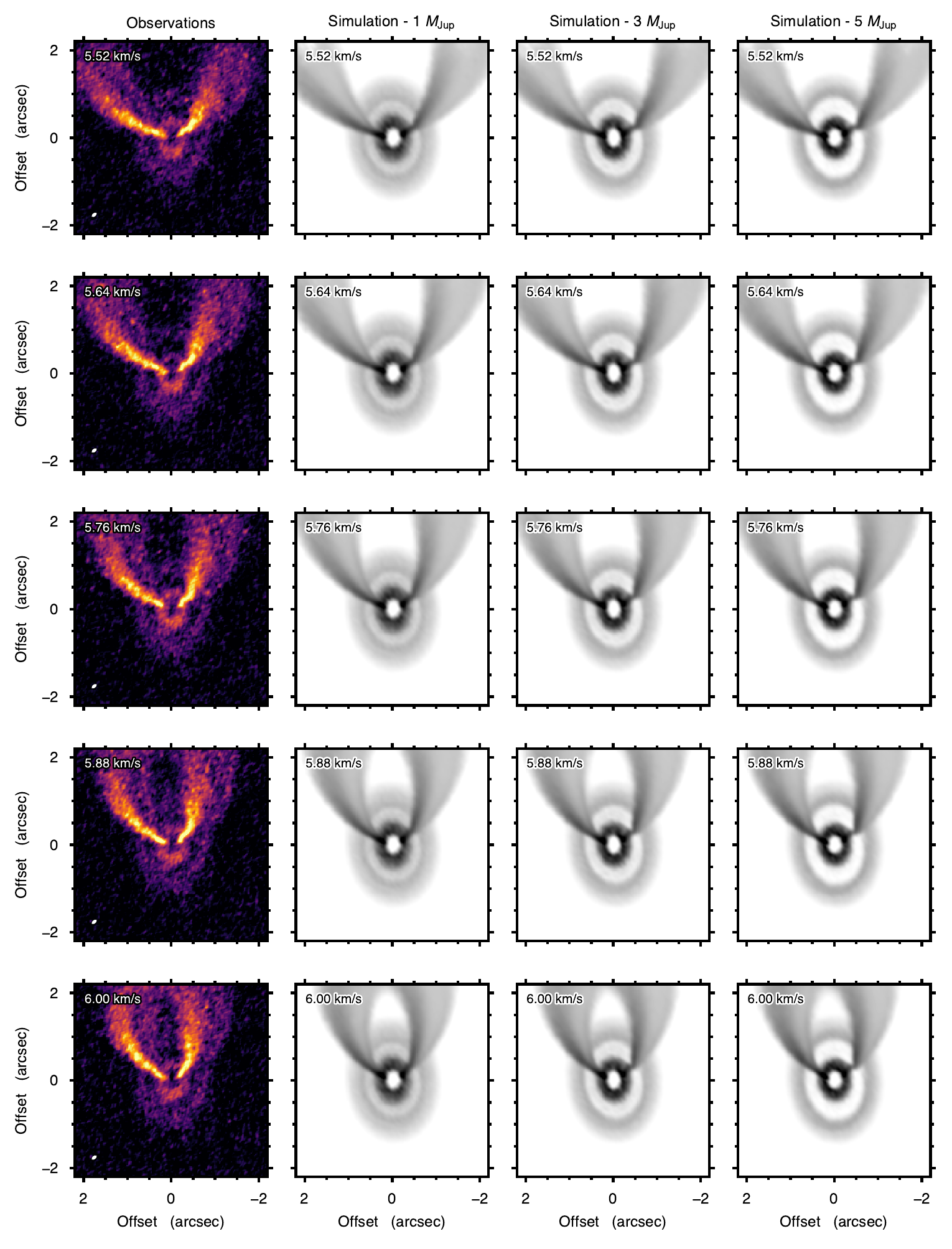}
    \caption{Comparison of the observed KPS in $^{13}$CO from the disk around HD~97048, presented in \citet{Pinte_ea_2019} in the left column. The right three columns show post-processed 3D hydrodynamic simulations with different mass planets, marked at the top of each column, to demonstrate how the mass of the planet results in more or less significant perturbations observed in the channel maps. In addition to the stronger perturbation in the gas emission, the gap in the continuum becomes more prominent with higher planet mass. Figure modelled after Fig.~4 from \citet{Pinte_ea_2019}.}
    \label{fig:KPS_Pinte}
\end{figure*}

Even more direct is the localised velocity perturbation induced by the planet itself. Mock channel maps produced from 3D simulations by \citet{Perez2015}, as shown in Fig.~\ref{fig:KPS_Perez}, demonstrate a clear twist in the channel maps due to the perturbed velocity structure. The rapid adoption of embedded planets to explain the twists and breaks observed in high resolution ALMA images was driven mainly by the relative simplicity of the model: a giant planet embedded in a gaseous disk. Varying the mass of the planet, while holding the disk physical properties fixed, resulted in significantly different features (Fig.~\ref{fig:KPS_Perez} and Fig.~\ref{fig:KPS_Pinte}) allowing for a constraint on the mass of the embedded planet. Added to this, matching the induced gap and ring structures observed in the sub-mm continuum provide a secondary test of the inferred planet mass \citep[e.g.][]{Pinte_ea_2019}.

While observations of both the gas and the dust can be modelled by embedded planets, there are several issues which must be addressed in order for this to be more widely adopted as a new method for hunting for planets.

First, the precise origins of the kinematic features generated by the planet in channel maps are not yet well understood. \citet{Perez2015} assumed that the main observable feature would be the circumplanetary disk (CPD). The CPD shows the strongest deviation from the background Keplerian motion, but is too small to be spatially resolved in the observations. Their predicted maps (see Fig.~\ref{fig:KPS_Perez}) show the CPD as a bright point-like source in the channel maps, but the detections by \citet{Pinte_ea_2018b} are of the larger scale KPS in the emission contours (more readily seen in the left two rows of Fig.~\ref{fig:KPS_Perez}). This feature arises from a larger scale disturbance to the flow, most likely from the wake launched by the planet at the Lindblad resonances \citep{Goldreich_Tremaine1979, Ogilvie_Lubow2002, Rafikov2002}. Another possibility is that we are detecting motion in the `horseshoe orbits' which are interior to the planet gap but outside of the Hill radius. Although obtaining high spatial and spectral observations would certainly help clarify what is being observed, achieving the sensitivity necessary to image these subtle deviations remains a challenge even for state-of-the-art instruments like ALMA and this particular question is best tackled from the theoretical side.

Second, matching against a suite of vertically isothermal 3D simulations, customised to the disk model, is currently the only way to measure the planet mass \citep[e.g.][]{Pinte_ea_2018b, Pinte_ea_2019}. It is unclear whether the planet mass measured in this way depends on, or is degenerate with, the disk model. A deeper understanding of the precise origin of the kinematic features caused by the planet wake would be valuable here, and would ideally lead to a more direct measurement of the planet mass using analytic formulae \citep[such as the relationship between gap depth and width frequently used to infer planet masses capable of opening gaps; e.g.][]{Kanagawa_ea_2016}.

In particular, the distinction between which velocity components are traced is essential. Currently simulations have focused on the rotational velocity, $v_{\phi}$, however \citet{Pinte_ea_2019} showed that both radial and vertical flows, $v_r$ and $v_z$, are driven by an embedded planet. A particularly interesting point is that these velocity components have different radial and vertical dependencies, suggesting that the use of multiple gas traces, each tracing separate vertical regions in the disk, would provide tighter constraints on the mass of the planet driving these perturbations, but also likely some information on the physical structure of the disk. The adoption of realistic temperature and density structures coupled with an accurate implementation of the local thermodynamics will be essential in relating the strength of velocity perturbations to a planet mass.

Third, one cannot yet assess the statistical significance of detections made in individual channel maps. This is mainly because the detection of KPS to date has been performed by eye \citep{Pinte_ea_2018b, Pinte_ea_2019}. A more quantitative and automated procedure for detection of localised kinematic features is needed. Further to this, a detailed study is needed on the impact of various observing configurations to the sensitivity, particularly with regard to possible imaging artefacts (see Section~\ref{sec:imaging}), although the most obvious of these can already be ruled out (e.g.~continuum subtraction, which affects all channels rather than just a few; \citealt{Pinte_ea_2019}). While this issue is primarily associated with the analysis of observations, a suite of `expected' emission morphologies spanning a range of disk and planet properties is required for the development of such methods.

\subsection{Meridional Flows}
\label{sec:theory:meridional_flows}

It was quickly established from early 3D simulations that, when a planet is massive enough to open a gap in the gas density, typically when the planet has reached the giant planet regime, a large vertical flow at the radius of the gap will be generated as the gas moves to maintain hydrostatic equilibrium in the gap \citep{Kley_ea_2001, Crida2006, Crida_ea_2009}. Such work prompted further exploration into the flow structure associated with an embedded planet, and in particular the way in which material is accreted onto the growing planet.

\citet{Tanigawa2012} used high resolution shearing-box simulations to demonstrate a complex system of flows around the embedded planet, noting in particular that the majority of the gas (up to $\sim 90\%$) is delivered through vertical flows onto the circumplanetary disks, rather than radially through the midplane. \citet{Szulagyi2014} and \citet{Morbidelli_ea_2014} further showed that this process drives meridional circulation: after falling towards the disk midplane, gas is driven radially away from the planet before returning to the surface in order to maintain hydrostatic equilibrium, a similar finding to \citet{Fung_Chiang2016}. The vertical motions were found to be strongest at the azimuthal location of the planet, extending between 60$^{\circ}$ and 120$^{\circ}$ behind the planet, although some vertical motion was found at the edge of the gap across the entire azimuthal extent of the disk \citep{Fung_Chiang2016}.

Using ALMA observations of the disk around HD~163296, \citet{Teague_ea_2019a} observed gas flows comparable in morphology and velocity as predicted for meridional flows, attributing them to three previously detected embedded planets \citep{Teague_ea_2018a, Pinte_ea_2018b}. However, as the observations used $^{12}$CO emission which probes only atmospheric regions ($z/r \sim 0.3$), a full characterisation of the flows was not possible. Future observations will aim to trace these flows using less abundant molecules, such as the less abundant CO isotopologues, $^{13}$CO and C$^{18}$O, allowing for an unambiguous detection of an embedded planet.

Although current simulations all qualitatively agree with one another, there are still several aspects which can be improved, particularly if the detection and characterisation of meridional flows is to be used to constrain the mass of the embedded planet. \citet{Szulagyi2014} and \citet{Morbidelli_ea_2014} argued that the meridional circulation is governed by viscous timescale (rather than the dynamical timescale, which is much shorter), and so the velocity of the in-falling gas is proportional to the depth of the gap.

Including realistic gas density and temperature structures, obtained through observations of multiple molecular lines, would allow for a more direct comparison between the flow velocities and morphologies. Furthermore, it is essential to understand how these flows interact in the case of more than one planet embedded within the disk. This was touched upon in the simulations presented in \citet{Teague_ea_2019a}, however absolutely requires additional work. Finally, synthetic observations of different molecular tracers are essential in designing the next wave of observations. For example, understanding if the vortex-like flows presented in \citet{Szulagyi2020} can be spatially resolved, or results in only a non-thermal broadening component will help guide observational design.

\subsection{Non-Planetary Scenarios}

While the observational evidence is beginning to point towards embedded planets as the culprit for producing gaps and rings in protoplanetary disks, there are a number of other mechanisms put forth in the literature able to reproduce the observed substructures seen in the dust. Due to the scarcity of high resolution observations of molecular line emission, nearly all work thus far has, by necessity, focused on the reproduction of the substructures observed in the dust. However, these mechanisms will undoubtedly drive localised velocity perturbations which can manifest as features similar to those discussed above and it is therefore essential to make predictions about kinematical features from these mechanisms that would allow us to distinguish between planet-related or other structures.

Broadly speaking, it is possible to split mechanisms into those which are associated with large scale flows or significant perturbations in the kinematical structure of the disk, and those that are not. Those in the latter category, such ice-line related structures \citep[e.g.][]{zhang2015} or secular gravitational instabilities \citep{Ward1976, ward2000}, can readily be ruled out by the large kinematical perturbations detected \citep{Pinte_ea_2018b, Pinte_ea_2019, Casassus_Perez_2019, Teague_ea_2019a}.

From those that do involve large dynamical flows, such as MHD zonal flows \citep[e.g.][]{johansen2009, dzy2010, uribe2011, simon2012, lyra2008, simon2014, bai2014, bai2015, bethune2017, suriano2017, simon2018, Suriano_ea_2018, riols2019}, both with and without the presence of a vertical magnetic field, the break down of axis-symmetric rings into localised vorticies via the Rossby-Wave Instability \citep{lovelace1999, lyra2009, lyra2015, Meheut_ea_2010, Meheut_ea_2012b, Meheut_ea_2012a, Surville_ea_2016} or the vertical shear instability \citep[e.g.][]{Nelson_ea_2013, Lin_Youdin_2015} or gravitational instability \citep[e.g.][]{Evans_ea_2019}, it is essential that predictions are made for how these manifest in the observed velocity maps. We note that some work has been done in the context of searches for turbulence with non-thermal line broadening \citep[e.g.][]{simon2015, Flock_ea_2017}, however these should be extended to consider larger flows which can be spatially resolved.

In sum, such theoretical work is essential if we are to correctly interpret the observations, both for planet-related and non-planet-related mechanisms. More accurate simulations of the dynamical structures associated with planet-disk interactions will help observers map the morphology of the observed spiral wakes to properties of the embedded planet, and design highly optimised observations to search for embedded planets. In particular, we advocate for:

\begin{enumerate}
    \item Focusing on 3D simulations as the flow structures we expect are inherently three dimensional in nature.
    \item Including thermodynamics within the simulations as the temperature strongly affecst the resulting velocities (higher temperatures result in higher gas pressure, in turn dampening gas velocities).
    \item Using as high resolution as possible in order to resolve the physics near hte planet. Too low resolutions will underestimate the torques which influence the gap size and the strength of the spiral wakes.
\end{enumerate}

In tandem, a gallery of non-planet related instabilities will allow for non-detections and false-positives to be more readily removed and help in the development of image analysis techniques focused on the extraction of kinematical information from astronomical images.

\section{Conclusions}
\label{sec:conclusions}

It is without question that ALMA has revolutionised our view of the planet formation environment and challenged our understanding of the planet formation process. The gaps, rings and spirals routinely observed in the sub-mm continuum, coupled with the comparable level of substructure seen in the scattered NIR light, strongly suggest that we are witnessing planet formation in action. Although observations of the dust in protoplanetary disk, unlike the probes of the gas, are able to achieve unparalleled spatial resolutions, they lack the ability to map out the kinematical structure, essential context to understand the origins of these structures.

Over the last couple of years, it has been shown that it is possible to observe small, percent-level deviations in the velocity structure of protoplanetary disks. The coincidence of these features with structures observed in the dust is highly indicative that they are due to embedded planets; providing an entirely unique probe of planet formation and detecting planets.

In this paper, we have presented the current state of the field with regard to the kinematical features associated with planet-disk interactions. We have demonstrated that contemporary models predict planets to drive significant deviations relative to the background near-Keplerian rotation of the disk, such as spiral wakes or large, vertical flows. We further describe the current suite of analysis techniques which have been adopted in order isolate these features in interferometric maps of molecular line emission, demonstrating that, while this field is in its infancy, there is much fertile ground for exploration.

Further to the methods used to analyse data, we highlight caveats and assumptions which must be made in order to extract these values. In particular, caution must be taken when interpreting images synthesised from interferometers due to the non-linear imaging process. We propose a series of best practices to mitigate these issues and a checklist of features which we believe constitute a kinematical detection of an embedded planet.

In addition, we discuss the current state of theoretical works looking at planet-disk interactions. While a coherent picture of the flow structures excited by and associated with an embedded planet appears to have been reached, there is much work to be done on understanding the details of these flows: how they change with different assumptions on the equation of state or the inclusion of realistic thermodynamics, for example. Continued work in this field is essential in aiding the interpretation of observations, in particular the what the features associated with planets should look like, and how they can be distinguished from non-planet related structures.

The work summarised here points to a bright future; we believe that ALMA is poised and ready to realise its full planet-hunting potential. Recent results have only scratched the surface of what is possible with this instrument. Bringing together novel image synthesis techniques and methods to extract key kinematical information from data, supported by a comprehensive foundation of theoretical work and simulations, we can begin to truly understand the earliest stages of planet formation.


\begin{acknowledgements}
This review was initiated during a workshop at the Center for Computational Astrophysics, which is part of the Flatiron Institute, supported by the Simons Foundation. We thank the staff for providing a stimulating and collaborative environment for these discussions. This paper makes use of the following ALMA data: 2016.1.00629.S, 2016.1.00484.L, 2015.1.00686.S, 2013.1.00366.S, and 2013.1.00601.S. ALMA is a partnership of ESO (representing its member states), NSF (USA) and NINS (Japan), together with NRC (Canada), NSC and ASIAA (Taiwan), and KASI (Republic of Korea), in cooperation with the Republic of Chile. The Joint ALMA Observatory is operated by ESO, AUI/NRAO and NAOJ.
JB acknowledges support by NASA through the NASA Hubble Fellowship grant \#HST-HF2-51427.001-A awarded  by  the  Space  Telescope  Science  Institute,  which  is  operated  by  the  Association  of  Universities  for  Research  in  Astronomy, Incorporated, under NASA contract NAS5-26555.
IC was supported by NASA through the NASA Hubble Fellowship grant HST-HF2-51405.001-A awarded by the Space Telescope Science Institute, which is operated by the Association of Universities for Research in Astronomy, Inc., for NASA, under contract NAS5-26555.
CH is a Winton Fellow and this work has been supported by Winton Philanthropies / The David and Claudia Harding Foundation.
SF acknowledges an ESO Fellowship. C.P. and D.J.P. acknowledge funding from the Australian Research Council via
FT170100040, FT130100034, and DP180104235.
JDI acknowledges support from the Science and Technology Facilities Council of the United Kingdom (STFC) under ST/R000549/1 and ST/T000287/1.
S.P acknowledges support from ANID-FONDECYT grant 1191934 and from the Joint Committee of ESO and Chile.
K.R.S. acknowledges the support of NASA through Hubble Fellowship grant HST-HF2-51419.001 awarded by the Space Telescope ScienceInstitute, which is operated by the Association of Universities for Research in Astronomy, Inc., for NASA, under contract NAS5-26555.
RT acknowledges support from the Smithsonian Institution as a Submillimeter Array (SMA) Fellow.
\end{acknowledgements}


\bibliography{bibliography}{}

\begin{thebibliography}{}
\makeatletter
\relax
\def\mn@urlcharsother{\let\do\@makeother \do\$\do\&\do\#\do\^\do\_\do\%\do\~}
\definecolor{darkblue}{rgb}{0,0,0.597656}
\def\mndoi{\begingroup\mn@urlcharsother \@ifnextchar [ {\mndoi@} {\mndoi@[]}}
\def\mndoi@[#1]#2{\def\@tempa{#1}\ifx\@tempa\@empty \href
  {http://dx.doi.org/#2} {\textcolor{darkblue}{doi:#2}}\else \href
  {http://dx.doi.org/#2} {\textcolor{darkblue}{#1}}\fi \endgroup}
\def\mn@eprint#1#2{\mn@eprint@#1:#2::\@nil}
\def\mn@eprint@arXiv#1{\href {http://arxiv.org/abs/#1} {{\tt arXiv:#1}}}
\def\mn@eprint@dblp#1{\href {http://dblp.uni-trier.de/rec/bibtex/#1.xml}
  {dblp:#1}}
\def\mn@eprint@#1:#2:#3:#4\@nil{\def\@tempa {#1}\def\@tempb {#2}\def\@tempc
  {#3}\ifx \@tempc \@empty \let \@tempc \@tempb \let \@tempb \@tempa \fi \ifx
  \@tempb \@empty \def\@tempb {arXiv}\fi \@ifundefined
  {mn@eprint@\@tempb}{\@tempb:\@tempc}{\expandafter \expandafter \csname
  mn@eprint@\@tempb\endcsname \expandafter{\@tempc}}}

\bibitem[\protect\citeauthoryear{{ALMA Partnership} et~al.,}{{ALMA Partnership}
  et~al.}{2015}]{ALMA_ea_2015}
{ALMA Partnership} et~al., 2015, \mndoi [\apjl] {10.1088/2041-8205/808/1/L3},
  \href {https://ui.adsabs.harvard.edu/abs/2015ApJ...808L...3A} {808, L3}

\bibitem[\protect\citeauthoryear{{Andrews} et~al.,}{{Andrews}
  et~al.}{2016}]{Andrews_ea_2016}
{Andrews} S.~M.,  et~al., 2016, \mndoi [\apjl] {10.3847/2041-8205/820/2/L40},
  \href {https://ui.adsabs.harvard.edu/abs/2016ApJ...820L..40A} {820, L40}

\bibitem[\protect\citeauthoryear{{Andrews} et~al.,}{{Andrews}
  et~al.}{2018}]{Andrews_ea_2018}
{Andrews} S.~M.,  et~al., 2018, \mndoi [\apjl] {10.3847/2041-8213/aaf741},
  \href {https://ui.adsabs.harvard.edu/abs/2018ApJ...869L..41A} {869, L41}

\bibitem[\protect\citeauthoryear{{Aoyama}, {Ikoma}  \& {Tanigawa}}{{Aoyama}
  et~al.}{2018}]{Aoyama2018}
{Aoyama} Y.,  {Ikoma} M.,   {Tanigawa} T.,  2018, \mndoi [\apj]
  {10.3847/1538-4357/aadc11}, \href
  {https://ui.adsabs.harvard.edu/abs/2018ApJ...866...84A} {866, 84}

\bibitem[\protect\citeauthoryear{{Bae}, {Pinilla}  \& {Birnstiel}}{{Bae}
  et~al.}{2018}]{Bae_ea_2018}
{Bae} J.,  {Pinilla} P.,   {Birnstiel} T.,  2018, \mndoi [\apjl]
  {10.3847/2041-8213/aadd51}, \href
  {https://ui.adsabs.harvard.edu/abs/2018ApJ...864L..26B} {864, L26}

\bibitem[\protect\citeauthoryear{{Bai}}{{Bai}}{2015}]{bai2015}
{Bai} X.-N.,  2015, \mndoi [\apj] {10.1088/0004-637X/798/2/84}, \href
  {https://ui.adsabs.harvard.edu/abs/2015ApJ...798...84B} {798, 84}

\bibitem[\protect\citeauthoryear{{Bai} \& {Stone}}{{Bai} \&
  {Stone}}{2014}]{bai2014}
{Bai} X.-N.,  {Stone} J.~M.,  2014, \mndoi [\apj] {10.1088/0004-637X/796/1/31},
  \href {https://ui.adsabs.harvard.edu/abs/2014ApJ...796...31B} {796, 31}

\bibitem[\protect\citeauthoryear{{Baruteau} et~al.,}{{Baruteau}
  et~al.}{2014}]{Baruteau2014}
{Baruteau} C.,  et~al., 2014, in {Beuther} H.,  {Klessen} R.~S.,  {Dullemond}
  C.~P.,   {Henning} T.,  eds, Protostars and Planets VI. p.~667 (\mn@eprint
  {arXiv} {1312.4293}), \mndoi{10.2458/azu_uapress_9780816531240-ch029}

\bibitem[\protect\citeauthoryear{{Benisty} et~al.,}{{Benisty}
  et~al.}{2015}]{Benisty2015}
{Benisty} M.,  et~al., 2015, \mndoi [\aap] {10.1051/0004-6361/201526011}, \href
  {https://ui.adsabs.harvard.edu/abs/2015A&A...578L...6B} {578, L6}

\bibitem[\protect\citeauthoryear{{Benz}, {Ida}, {Alibert}, {Lin}  \&
  {Mordasini}}{{Benz} et~al.}{2014}]{Benz2014}
{Benz} W.,  {Ida} S.,  {Alibert} Y.,  {Lin} D.,   {Mordasini} C.,  2014, in
  {Beuther} H.,  {Klessen} R.~S.,  {Dullemond} C.~P.,   {Henning} T.,  eds,
  Protostars and Planets VI. p.~691 (\mn@eprint {arXiv} {1402.7086}),
  \mndoi{10.2458/azu_uapress_9780816531240-ch030}

\bibitem[\protect\citeauthoryear{{Bergin}, {Du}, {Cleeves}, {Blake}, {Schwarz},
  {Visser}  \& {Zhang}}{{Bergin} et~al.}{2016}]{Bergin_ea_2016}
{Bergin} E.~A.,  {Du} F.,  {Cleeves} L.~I.,  {Blake} G.~A.,  {Schwarz} K.,
  {Visser} R.,   {Zhang} K.,  2016, \mndoi [\apj]
  {10.3847/0004-637X/831/1/101}, \href
  {https://ui.adsabs.harvard.edu/abs/2016ApJ...831..101B} {831, 101}

\bibitem[\protect\citeauthoryear{{B{\'e}thune}, {Lesur}  \&
  {Ferreira}}{{B{\'e}thune} et~al.}{2017}]{bethune2017}
{B{\'e}thune} W.,  {Lesur} G.,   {Ferreira} J.,  2017, \mndoi [\aap]
  {10.1051/0004-6361/201630056}, \href
  {https://ui.adsabs.harvard.edu/abs/2017A&A...600A..75B} {600, A75}

\bibitem[\protect\citeauthoryear{{Booth}, {Clarke}, {Madhusudhan}  \&
  {Ilee}}{{Booth} et~al.}{2017}]{Booth2017}
{Booth} R.~A.,  {Clarke} C.~J.,  {Madhusudhan} N.,   {Ilee} J.~D.,  2017,
  \mndoi [\mnras] {10.1093/mnras/stx1103}, \href
  {https://ui.adsabs.harvard.edu/abs/2017MNRAS.469.3994B} {469, 3994}

\bibitem[\protect\citeauthoryear{{Bosman}, {Cridland}  \& {Miguel}}{{Bosman}
  et~al.}{2019}]{Bosman_ea_2019}
{Bosman} A.~D.,  {Cridland} A.~J.,   {Miguel} Y.,  2019, \mndoi [\aap]
  {10.1051/0004-6361/201936827}, \href
  {https://ui.adsabs.harvard.edu/abs/2019A&A...632L..11B} {632, L11}

\bibitem[\protect\citeauthoryear{{Boss}}{{Boss}}{1997}]{Boss_1997}
{Boss} A.~P.,  1997, \mndoi [Science] {10.1126/science.276.5320.1836}, \href
  {https://ui.adsabs.harvard.edu/abs/1997Sci...276.1836B} {276, 1836}

\bibitem[\protect\citeauthoryear{{Bryden}, {Chen}, {Lin}, {Nelson}  \&
  {Papaloizou}}{{Bryden} et~al.}{1999}]{Bryden_ea_1999}
{Bryden} G.,  {Chen} X.,  {Lin} D.~N.~C.,  {Nelson} R.~P.,   {Papaloizou} J.
  C.~B.,  1999, \mndoi [\apj] {10.1086/306917}, \href
  {https://ui.adsabs.harvard.edu/abs/1999ApJ...514..344B} {514, 344}

\bibitem[\protect\citeauthoryear{{Cadman}, {Rice}, {Hall}, {Haworth}  \&
  {Biller}}{{Cadman} et~al.}{2020}]{cadman2020}
{Cadman} J.,  {Rice} K.,  {Hall} C.,  {Haworth} T.~J.,   {Biller} B.,  2020,
  \mndoi [\mnras] {10.1093/mnras/staa187}, \href
  {https://ui.adsabs.harvard.edu/abs/2020MNRAS.492.5041C} {492, 5041}

\bibitem[\protect\citeauthoryear{{C{\'a}rcamo}, {Rom{\'a}n}, {Casassus},
  {Moral}  \& {Rannou}}{{C{\'a}rcamo} et~al.}{2018}]{carcamo2018}
{C{\'a}rcamo} M.,  {Rom{\'a}n} P.~E.,  {Casassus} S.,  {Moral} V.,   {Rannou}
  F.~R.,  2018, \mndoi [Astronomy and Computing] {10.1016/j.ascom.2017.11.003},
  \href {https://ui.adsabs.harvard.edu/abs/2018A&C....22...16C} {22, 16}

\bibitem[\protect\citeauthoryear{{Casassus} \& {P{\'e}rez}}{{Casassus} \&
  {P{\'e}rez}}{2019}]{Casassus_Perez_2019}
{Casassus} S.,  {P{\'e}rez} S.,  2019, \mndoi [\apjl]
  {10.3847/2041-8213/ab4425}, \href
  {https://ui.adsabs.harvard.edu/abs/2019ApJ...883L..41C} {883, L41}

\bibitem[\protect\citeauthoryear{{Cazzoletti}, {van Dishoeck}, {Visser},
  {Facchini}  \& {Bruderer}}{{Cazzoletti} et~al.}{2018}]{Cazzoletti_ea_2018}
{Cazzoletti} P.,  {van Dishoeck} E.~F.,  {Visser} R.,  {Facchini} S.,
  {Bruderer} S.,  2018, \mndoi [\aap] {10.1051/0004-6361/201731457}, \href
  {https://ui.adsabs.harvard.edu/abs/2018A&A...609A..93C} {609, A93}

\bibitem[\protect\citeauthoryear{{Clarke}}{{Clarke}}{2009}]{Clarke2009}
{Clarke} C.~J.,  2009, \mndoi [\mnras] {10.1111/j.1365-2966.2009.14774.x},
  \href {https://ui.adsabs.harvard.edu/abs/2009MNRAS.396.1066C} {396, 1066}

\bibitem[\protect\citeauthoryear{{Cleeves}, {{\"O}berg}, {Wilner}, {Huang},
  {Loomis}, {Andrews}  \& {Guzman}}{{Cleeves} et~al.}{2018}]{Cleeves2018}
{Cleeves} L.~I.,  {{\"O}berg} K.~I.,  {Wilner} D.~J.,  {Huang} J.,  {Loomis}
  R.~A.,  {Andrews} S.~M.,   {Guzman} V.~V.,  2018, \mndoi [\apj]
  {10.3847/1538-4357/aade96}, \href
  {https://ui.adsabs.harvard.edu/abs/2018ApJ...865..155C} {865, 155}

\bibitem[\protect\citeauthoryear{{Cornwell}}{{Cornwell}}{2008}]{cornwell2008}
{Cornwell} T.~J.,  2008, \mndoi [IEEE Journal of Selected Topics in Signal
  Processing] {10.1109/JSTSP.2008.2006388}, \href
  {https://ui.adsabs.harvard.edu/abs/2008ISTSP...2..793C} {2, 793}

\bibitem[\protect\citeauthoryear{{Cornwell} \& {Evans}}{{Cornwell} \&
  {Evans}}{1985}]{cornwell1985}
{Cornwell} T.~J.,  {Evans} K.~F.,  1985, \aap, \href
  {https://ui.adsabs.harvard.edu/abs/1985A&A...143...77C} {143, 77}

\bibitem[\protect\citeauthoryear{{Crida}, {Morbidelli}  \& {Masset}}{{Crida}
  et~al.}{2006}]{Crida2006}
{Crida} A.,  {Morbidelli} A.,   {Masset} F.,  2006, \mndoi [Icarus]
  {10.1016/j.icarus.2005.10.007}, \href
  {https://ui.adsabs.harvard.edu/abs/2006Icar..181..587C} {181, 587}

\bibitem[\protect\citeauthoryear{{Crida}, {Baruteau}, {Kley}  \&
  {Masset}}{{Crida} et~al.}{2009}]{Crida_ea_2009}
{Crida} A.,  {Baruteau} C.,  {Kley} W.,   {Masset} F.,  2009, \mndoi [\aap]
  {10.1051/0004-6361/200811608}, \href
  {https://ui.adsabs.harvard.edu/abs/2009A&A...502..679C} {502, 679}

\bibitem[\protect\citeauthoryear{{Cridland}, {Pudritz}  \& {Alessi}}{{Cridland}
  et~al.}{2016}]{Cridland_ea_2016}
{Cridland} A.~J.,  {Pudritz} R.~E.,   {Alessi} M.,  2016, \mndoi [\mnras]
  {10.1093/mnras/stw1511}, \href
  {https://ui.adsabs.harvard.edu/abs/2016MNRAS.461.3274C} {461, 3274}

\bibitem[\protect\citeauthoryear{{Cridland}, {Pudritz}, {Birnstiel}, {Cleeves}
  \& {Bergin}}{{Cridland} et~al.}{2017}]{Cridland_ea_2017}
{Cridland} A.~J.,  {Pudritz} R.~E.,  {Birnstiel} T.,  {Cleeves} L.~I.,
  {Bergin} E.~A.,  2017, \mndoi [\mnras] {10.1093/mnras/stx1069}, \href
  {https://ui.adsabs.harvard.edu/abs/2017MNRAS.469.3910C} {469, 3910}

\bibitem[\protect\citeauthoryear{{Cridland}, {Pudritz}  \& {Alessi}}{{Cridland}
  et~al.}{2019}]{cridland19}
{Cridland} A.~J.,  {Pudritz} R.~E.,   {Alessi} M.,  2019, \mndoi [\mnras]
  {10.1093/mnras/stz008}, \href
  {https://ui.adsabs.harvard.edu/abs/2019MNRAS.484..345C} {484, 345}

\bibitem[\protect\citeauthoryear{{Cugno} et~al.,}{{Cugno}
  et~al.}{2019}]{Cugno2019}
{Cugno} G.,  et~al., 2019, \mndoi [\aap] {10.1051/0004-6361/201834170}, \href
  {https://ui.adsabs.harvard.edu/abs/2019A&A...622A.156C} {622, A156}

\bibitem[\protect\citeauthoryear{{Czekala} \& {Loomis}}{{Czekala} \&
  {Loomis}}{2020}]{mpol}
{Czekala} I.,  {Loomis} R.,  2020, iancze/MPoL: Pip installable package,
  \mndoi{10.5281/zenodo.3647603}, \url {https://doi.org/10.5281/zenodo.3647603}

\bibitem[\protect\citeauthoryear{{Czekala} et~al.,}{{Czekala}
  et~al.}{2017}]{czekala2017b}
{Czekala} I.,  et~al., 2017, \mndoi [\apj] {10.3847/1538-4357/aa9be7}, \href
  {https://ui.adsabs.harvard.edu/abs/2017ApJ...851..132C} {851, 132}

\bibitem[\protect\citeauthoryear{{Dipierro} \& {Laibe}}{{Dipierro} \&
  {Laibe}}{2017}]{Dipierro_Laibe2017}
{Dipierro} G.,  {Laibe} G.,  2017, \mndoi [\mnras] {10.1093/mnras/stx977},
  \href {https://ui.adsabs.harvard.edu/abs/2017MNRAS.469.1932D} {469, 1932}

\bibitem[\protect\citeauthoryear{{Dipierro}, {Price}, {Laibe}, {Hirsh},
  {Cerioli}  \& {Lodato}}{{Dipierro} et~al.}{2015}]{Dipierro_ea_2015}
{Dipierro} G.,  {Price} D.,  {Laibe} G.,  {Hirsh} K.,  {Cerioli} A.,   {Lodato}
  G.,  2015, \mndoi [\mnras] {10.1093/mnrasl/slv105}, \href
  {https://ui.adsabs.harvard.edu/abs/2015MNRAS.453L..73D} {453, L73}

\bibitem[\protect\citeauthoryear{{Dipierro}, {Laibe}, {Price}  \&
  {Lodato}}{{Dipierro} et~al.}{2016}]{Dipierro_ea_2016}
{Dipierro} G.,  {Laibe} G.,  {Price} D.~J.,   {Lodato} G.,  2016, \mndoi
  [\mnras] {10.1093/mnrasl/slw032}, \href
  {https://ui.adsabs.harvard.edu/abs/2016MNRAS.459L...1D} {459, L1}

\bibitem[\protect\citeauthoryear{{Dong} \& {Fung}}{{Dong} \&
  {Fung}}{2017}]{Dong2017a}
{Dong} R.,  {Fung} J.,  2017, \mndoi [\apj] {10.3847/1538-4357/835/2/146},
  \href {https://ui.adsabs.harvard.edu/abs/2017ApJ...835..146D} {835, 146}

\bibitem[\protect\citeauthoryear{{Dong}, {Li}, {Chiang}  \& {Li}}{{Dong}
  et~al.}{2017}]{Dong2017b}
{Dong} R.,  {Li} S.,  {Chiang} E.,   {Li} H.,  2017, \mndoi [\apj]
  {10.3847/1538-4357/aa72f2}, \href
  {https://ui.adsabs.harvard.edu/abs/2017ApJ...843..127D} {843, 127}

\bibitem[\protect\citeauthoryear{{Dong}, {Liu}  \& {Fung}}{{Dong}
  et~al.}{2019}]{Dong2019}
{Dong} R.,  {Liu} S.-Y.,   {Fung} J.,  2019, \mndoi [\apj]
  {10.3847/1538-4357/aaf38e}, \href
  {https://ui.adsabs.harvard.edu/abs/2019ApJ...870...72D} {870, 72}

\bibitem[\protect\citeauthoryear{{Duffell} \& {MacFadyen}}{{Duffell} \&
  {MacFadyen}}{2013}]{Duffell2013}
{Duffell} P.~C.,  {MacFadyen} A.~I.,  2013, \mndoi [\apj]
  {10.1088/0004-637X/769/1/41}, \href
  {https://ui.adsabs.harvard.edu/abs/2013ApJ...769...41D} {769, 41}

\bibitem[\protect\citeauthoryear{{Dullemond} et~al.,}{{Dullemond}
  et~al.}{2018}]{dullemond2018b}
{Dullemond} C.~P.,  et~al., 2018, \mndoi [\apjl] {10.3847/2041-8213/aaf742},
  \href {https://ui.adsabs.harvard.edu/abs/2018ApJ...869L..46D} {869, L46}

\bibitem[\protect\citeauthoryear{{Dzyurkevich}, {Flock}, {Turner}, {Klahr}  \&
  {Henning}}{{Dzyurkevich} et~al.}{2010}]{dzy2010}
{Dzyurkevich} N.,  {Flock} M.,  {Turner} N.~J.,  {Klahr} H.,   {Henning} T.,
  2010, \mndoi [\aap] {10.1051/0004-6361/200912834}, \href
  {https://ui.adsabs.harvard.edu/abs/2010A&A...515A..70D} {515, A70}

\bibitem[\protect\citeauthoryear{{Evans}, {Hartquist}, {Caselli}, {Boley},
  {Ilee}  \& {Rawlings}}{{Evans} et~al.}{2019}]{Evans_ea_2019}
{Evans} M.~G.,  {Hartquist} T.~W.,  {Caselli} P.,  {Boley} A.~C.,  {Ilee}
  J.~D.,   {Rawlings} J.~M.~C.,  2019, \mndoi [\mnras] {10.1093/mnras/sty2765},
  \href {https://ui.adsabs.harvard.edu/abs/2019MNRAS.483.1266E} {483, 1266}

\bibitem[\protect\citeauthoryear{{Event Horizon Telescope Collaboration}
  et~al.,}{{Event Horizon Telescope Collaboration} et~al.}{2019}]{eht-IV-2019}
{Event Horizon Telescope Collaboration} et~al., 2019, \mndoi [\apjl]
  {10.3847/2041-8213/ab0e85}, \href
  {https://ui.adsabs.harvard.edu/abs/2019ApJ...875L...4E} {875, L4}

\bibitem[\protect\citeauthoryear{{Facchini}, {Pinilla}, {van Dishoeck}  \& {de
  Juan Ovelar}}{{Facchini} et~al.}{2018}]{Facchini_ea_2018}
{Facchini} S.,  {Pinilla} P.,  {van Dishoeck} E.~F.,   {de Juan Ovelar} M.,
  2018, \mndoi [\aap] {10.1051/0004-6361/201731390}, \href
  {https://ui.adsabs.harvard.edu/abs/2018A&A...612A.104F} {612, A104}

\bibitem[\protect\citeauthoryear{{Facchini} et~al.,}{{Facchini}
  et~al.}{2020}]{Facchini_ea_2020}
{Facchini} S.,  et~al., 2020, \mndoi [\aap] {10.1051/0004-6361/202038027},
  \href {https://ui.adsabs.harvard.edu/abs/2020A&A...639A.121F} {639, A121}

\bibitem[\protect\citeauthoryear{{Flaherty}, {Hughes}, {Rosenfeld}, {Andrews},
  {Chiang}, {Simon}, {Kerzner}  \& {Wilner}}{{Flaherty}
  et~al.}{2015}]{Flaherty2015}
{Flaherty} K.~M.,  {Hughes} A.~M.,  {Rosenfeld} K.~A.,  {Andrews} S.~M.,
  {Chiang} E.,  {Simon} J.~B.,  {Kerzner} S.,   {Wilner} D.~J.,  2015, \mndoi
  [\apj] {10.1088/0004-637X/813/2/99}, \href
  {https://ui.adsabs.harvard.edu/abs/2015ApJ...813...99F} {813, 99}

\bibitem[\protect\citeauthoryear{{Flaherty} et~al.,}{{Flaherty}
  et~al.}{2017}]{Flaherty2017}
{Flaherty} K.~M.,  et~al., 2017, \mndoi [\apj] {10.3847/1538-4357/aa79f9},
  \href {https://ui.adsabs.harvard.edu/abs/2017ApJ...843..150F} {843, 150}

\bibitem[\protect\citeauthoryear{{Flock}, {Ruge}, {Dzyurkevich}, {Henning},
  {Klahr}  \& {Wolf}}{{Flock} et~al.}{2015}]{flock_2015}
{Flock} M.,  {Ruge} J.~P.,  {Dzyurkevich} N.,  {Henning} T.,  {Klahr} H.,
  {Wolf} S.,  2015, \mndoi [\aap] {10.1051/0004-6361/201424693}, \href
  {https://ui.adsabs.harvard.edu/abs/2015A&A...574A..68F} {574, A68}

\bibitem[\protect\citeauthoryear{{Flock}, {Nelson}, {Turner}, {Bertrang},
  {Carrasco-Gonz{\'a}lez}, {Henning}, {Lyra}  \& {Teague}}{{Flock}
  et~al.}{2017}]{Flock_ea_2017}
{Flock} M.,  {Nelson} R.~P.,  {Turner} N.~J.,  {Bertrang} G. H.~M.,
  {Carrasco-Gonz{\'a}lez} C.,  {Henning} T.,  {Lyra} W.,   {Teague} R.,  2017,
  \mndoi [\apj] {10.3847/1538-4357/aa943f}, \href
  {https://ui.adsabs.harvard.edu/abs/2017ApJ...850..131F} {850, 131}

\bibitem[\protect\citeauthoryear{{Forgan} \& {Rice}}{{Forgan} \&
  {Rice}}{2011}]{forgan2011}
{Forgan} D.,  {Rice} K.,  2011, \mndoi [\mnras]
  {10.1111/j.1365-2966.2011.19380.x}, \href
  {https://ui.adsabs.harvard.edu/abs/2011MNRAS.417.1928F} {417, 1928}

\bibitem[\protect\citeauthoryear{{Forgan} \& {Rice}}{{Forgan} \&
  {Rice}}{2013}]{forgan2013}
{Forgan} D.,  {Rice} K.,  2013, \mndoi [\mnras] {10.1093/mnras/stt672}, \href
  {https://ui.adsabs.harvard.edu/abs/2013MNRAS.432.3168F} {432, 3168}

\bibitem[\protect\citeauthoryear{{Forgan}, {Hall}, {Meru}  \& {Rice}}{{Forgan}
  et~al.}{2018}]{forgan2018}
{Forgan} D.~H.,  {Hall} C.,  {Meru} F.,   {Rice} W.~K.~M.,  2018, \mndoi
  [\mnras] {10.1093/mnras/stx2870}, \href
  {https://ui.adsabs.harvard.edu/abs/2018MNRAS.474.5036F} {474, 5036}

\bibitem[\protect\citeauthoryear{{Fulton} \& {Petigura}}{{Fulton} \&
  {Petigura}}{2018}]{Fulton2018}
{Fulton} B.~J.,  {Petigura} E.~A.,  2018, \mndoi [\aj]
  {10.3847/1538-3881/aae828}, \href
  {https://ui.adsabs.harvard.edu/abs/2018AJ....156..264F} {156, 264}

\bibitem[\protect\citeauthoryear{{Fung} \& {Chiang}}{{Fung} \&
  {Chiang}}{2016}]{Fung_Chiang2016}
{Fung} J.,  {Chiang} E.,  2016, \mndoi [\apj] {10.3847/0004-637X/832/2/105},
  \href {https://ui.adsabs.harvard.edu/abs/2016ApJ...832..105F} {832, 105}

\bibitem[\protect\citeauthoryear{{Fung}, {Shi}  \& {Chiang}}{{Fung}
  et~al.}{2014}]{Fung2014}
{Fung} J.,  {Shi} J.-M.,   {Chiang} E.,  2014, \mndoi [\apj]
  {10.1088/0004-637X/782/2/88}, \href
  {https://ui.adsabs.harvard.edu/abs/2014ApJ...782...88F} {782, 88}

\bibitem[\protect\citeauthoryear{{Garufi} et~al.,}{{Garufi}
  et~al.}{2013}]{Garufi2013}
{Garufi} A.,  et~al., 2013, \mndoi [\aap] {10.1051/0004-6361/201322429}, \href
  {https://ui.adsabs.harvard.edu/abs/2013A&A...560A.105G} {560, A105}

\bibitem[\protect\citeauthoryear{{Goldreich} \& {Tremaine}}{{Goldreich} \&
  {Tremaine}}{1979}]{Goldreich_Tremaine1979}
{Goldreich} P.,  {Tremaine} S.,  1979, \mndoi [\apj] {10.1086/157448}, \href
  {https://ui.adsabs.harvard.edu/abs/1979ApJ...233..857G} {233, 857}

\bibitem[\protect\citeauthoryear{{Goldreich} \& {Tremaine}}{{Goldreich} \&
  {Tremaine}}{1980}]{Goldreich1980}
{Goldreich} P.,  {Tremaine} S.,  1980, \mndoi [\apj] {10.1086/158356}, \href
  {https://ui.adsabs.harvard.edu/abs/1980ApJ...241..425G} {241, 425}

\bibitem[\protect\citeauthoryear{{Guzm{\'a}n} et~al.,}{{Guzm{\'a}n}
  et~al.}{2018}]{guzman2018}
{Guzm{\'a}n} V.~V.,  et~al., 2018, \mndoi [\apjl] {10.3847/2041-8213/aaedae},
  \href {https://ui.adsabs.harvard.edu/abs/2018ApJ...869L..48G} {869, L48}

\bibitem[\protect\citeauthoryear{{Gyeol Yun}, {Kim}, {Bae}  \& {Han}}{{Gyeol
  Yun} et~al.}{2019}]{Yun_ea_2019}
{Gyeol Yun} H.,  {Kim} W.-T.,  {Bae} J.,   {Han} C.,  2019, \mndoi [\apj]
  {10.3847/1538-4357/ab3fab}, \href
  {https://ui.adsabs.harvard.edu/abs/2019ApJ...884..142G} {884, 142}

\bibitem[\protect\citeauthoryear{{Haffert}, {Bohn}, {de Boer}, {Snellen},
  {Brinchmann}, {Girard}, {Keller}  \& {Bacon}}{{Haffert}
  et~al.}{2019}]{Haffert_ea_2019}
{Haffert} S.~Y.,  {Bohn} A.~J.,  {de Boer} J.,  {Snellen} I.~A.~G.,
  {Brinchmann} J.,  {Girard} J.~H.,  {Keller} C.~U.,   {Bacon} R.,  2019,
  \mndoi [Nature Astronomy] {10.1038/s41550-019-0780-5}, \href
  {https://ui.adsabs.harvard.edu/abs/2019NatAs...3..749H} {3, 749}

\bibitem[\protect\citeauthoryear{{Hall}, {Forgan}  \& {Rice}}{{Hall}
  et~al.}{2017}]{hall2017}
{Hall} C.,  {Forgan} D.,   {Rice} K.,  2017, \mndoi [\mnras]
  {10.1093/mnras/stx1244}, \href
  {https://ui.adsabs.harvard.edu/abs/2017MNRAS.470.2517H} {470, 2517}

\bibitem[\protect\citeauthoryear{{Hashimoto} et~al.,}{{Hashimoto}
  et~al.}{2011}]{Hashimoto2011}
{Hashimoto} J.,  et~al., 2011, \mndoi [\apjl] {10.1088/2041-8205/729/2/L17},
  \href {https://ui.adsabs.harvard.edu/abs/2011ApJ...729L..17H} {729, L17}

\bibitem[\protect\citeauthoryear{{Hezaveh} et~al.,}{{Hezaveh}
  et~al.}{2013}]{hezaveh2013}
{Hezaveh} Y.~D.,  et~al., 2013, \mndoi [\apj] {10.1088/0004-637X/767/2/132},
  \href {https://ui.adsabs.harvard.edu/abs/2013ApJ...767..132H} {767, 132}

\bibitem[\protect\citeauthoryear{{H{\"o}gbom}}{{H{\"o}gbom}}{1974}]{hogbom1974}
{H{\"o}gbom} J.~A.,  1974, \aaps, \href
  {https://ui.adsabs.harvard.edu/abs/1974A&AS...15..417H} {15, 417}

\bibitem[\protect\citeauthoryear{{Hsu}, {Ford}, {Ragozzine}  \& {Ashby}}{{Hsu}
  et~al.}{2019}]{Hsu2019}
{Hsu} D.~C.,  {Ford} E.~B.,  {Ragozzine} D.,   {Ashby} K.,  2019, \mndoi [\aj]
  {10.3847/1538-3881/ab31ab}, \href
  {https://ui.adsabs.harvard.edu/abs/2019AJ....158..109H} {158, 109}

\bibitem[\protect\citeauthoryear{{Huang} et~al.,}{{Huang}
  et~al.}{2018}]{Huang_ea_2018}
{Huang} J.,  et~al., 2018, \mndoi [\apj] {10.3847/1538-4357/aaa1e7}, \href
  {https://ui.adsabs.harvard.edu/abs/2018ApJ...852..122H} {852, 122}

\bibitem[\protect\citeauthoryear{{Ilee} et~al.,}{{Ilee}
  et~al.}{2017}]{Ilee2017}
{Ilee} J.~D.,  et~al., 2017, \mndoi [\mnras] {10.1093/mnras/stx1966}, \href
  {https://ui.adsabs.harvard.edu/abs/2017MNRAS.472..189I} {472, 189}

\bibitem[\protect\citeauthoryear{{Isella} \& {Turner}}{{Isella} \&
  {Turner}}{2018}]{Isella_Turner_2018}
{Isella} A.,  {Turner} N.~J.,  2018, \mndoi [\apj] {10.3847/1538-4357/aabb07},
  \href {https://ui.adsabs.harvard.edu/abs/2018ApJ...860...27I} {860, 27}

\bibitem[\protect\citeauthoryear{{Isella} et~al.,}{{Isella}
  et~al.}{2016}]{Isella_ea_2016}
{Isella} A.,  et~al., 2016, \mndoi [\prl] {10.1103/PhysRevLett.117.251101},
  \href {https://ui.adsabs.harvard.edu/abs/2016PhRvL.117y1101I} {117, 251101}

\bibitem[\protect\citeauthoryear{{Isella} et~al.,}{{Isella}
  et~al.}{2018}]{Isella_ea_2018}
{Isella} A.,  et~al., 2018, \mndoi [\apjl] {10.3847/2041-8213/aaf747}, \href
  {https://ui.adsabs.harvard.edu/abs/2018ApJ...869L..49I} {869, L49}

\bibitem[\protect\citeauthoryear{{Isella}, {Benisty}, {Teague}, {Bae},
  {Keppler}, {Facchini}  \& {P{\'e}rez}}{{Isella}
  et~al.}{2019}]{Isella_ea_2019}
{Isella} A.,  {Benisty} M.,  {Teague} R.,  {Bae} J.,  {Keppler} M.,  {Facchini}
  S.,   {P{\'e}rez} L.,  2019, \mndoi [\apjl] {10.3847/2041-8213/ab2a12}, \href
  {https://ui.adsabs.harvard.edu/abs/2019ApJ...879L..25I} {879, L25}

\bibitem[\protect\citeauthoryear{{Johansen} \& {Lambrechts}}{{Johansen} \&
  {Lambrechts}}{2017}]{Johansen2017}
{Johansen} A.,  {Lambrechts} M.,  2017, \mndoi [Annual Review of Earth and
  Planetary Sciences] {10.1146/annurev-earth-063016-020226}, \href
  {https://ui.adsabs.harvard.edu/abs/2017AREPS..45..359J} {45, 359}

\bibitem[\protect\citeauthoryear{Johansen, Youdin  \& Klahr}{Johansen
  et~al.}{2009}]{johansen2009}
Johansen A.,  Youdin A.,   Klahr H.,  2009, The Astrophysical Journal, 697,
  1269

\bibitem[\protect\citeauthoryear{{Jorsater} \& {van Moorsel}}{{Jorsater} \&
  {van Moorsel}}{1995}]{Jorsater_vanMoorsel_1995}
{Jorsater} S.,  {van Moorsel} G.~A.,  1995, \mndoi [\aj] {10.1086/117668},
  \href {https://ui.adsabs.harvard.edu/abs/1995AJ....110.2037J} {110, 2037}

\bibitem[\protect\citeauthoryear{{Junklewitz}, {Bell}, {Selig}  \&
  {En{\ss}lin}}{{Junklewitz} et~al.}{2016}]{junklewitz2016}
{Junklewitz} H.,  {Bell} M.~R.,  {Selig} M.,   {En{\ss}lin} T.~A.,  2016,
  \mndoi [\aap] {10.1051/0004-6361/201323094}, \href
  {https://ui.adsabs.harvard.edu/abs/2016A&A...586A..76J} {586, A76}

\bibitem[\protect\citeauthoryear{{Kanagawa}, {Muto}, {Tanaka}, {Tanigawa},
  {Takeuchi}, {Tsukagoshi}  \& {Momose}}{{Kanagawa}
  et~al.}{2015}]{Kanagawa2015}
{Kanagawa} K.~D.,  {Muto} T.,  {Tanaka} H.,  {Tanigawa} T.,  {Takeuchi} T.,
  {Tsukagoshi} T.,   {Momose} M.,  2015, \mndoi [\apjl]
  {10.1088/2041-8205/806/1/L15}, \href
  {https://ui.adsabs.harvard.edu/abs/2015ApJ...806L..15K} {806, L15}

\bibitem[\protect\citeauthoryear{{Kanagawa}, {Muto}, {Tanaka}, {Tanigawa},
  {Takeuchi}, {Tsukagoshi}  \& {Momose}}{{Kanagawa}
  et~al.}{2016}]{Kanagawa_ea_2016}
{Kanagawa} K.~D.,  {Muto} T.,  {Tanaka} H.,  {Tanigawa} T.,  {Takeuchi} T.,
  {Tsukagoshi} T.,   {Momose} M.,  2016, \mndoi [\pasj] {10.1093/pasj/psw037},
  \href {https://ui.adsabs.harvard.edu/abs/2016PASJ...68...43K} {68, 43}

\bibitem[\protect\citeauthoryear{{Keppler} et~al.,}{{Keppler}
  et~al.}{2018}]{Keppler_ea_2018}
{Keppler} M.,  et~al., 2018, \mndoi [\aap] {10.1051/0004-6361/201832957}, \href
  {https://ui.adsabs.harvard.edu/abs/2018A&A...617A..44K} {617, A44}

\bibitem[\protect\citeauthoryear{{Keppler} et~al.,}{{Keppler}
  et~al.}{2019}]{Keppler_ea_2019}
{Keppler} M.,  et~al., 2019, \mndoi [\aap] {10.1051/0004-6361/201935034}, \href
  {https://ui.adsabs.harvard.edu/abs/2019A&A...625A.118K} {625, A118}

\bibitem[\protect\citeauthoryear{{Klahr} \& {Bodenheimer}}{{Klahr} \&
  {Bodenheimer}}{2003}]{Klahr_Bodenheimer_2003}
{Klahr} H.~H.,  {Bodenheimer} P.,  2003, \mndoi [\apj] {10.1086/344743}, \href
  {https://ui.adsabs.harvard.edu/abs/2003ApJ...582..869K} {582, 869}

\bibitem[\protect\citeauthoryear{{Kley}}{{Kley}}{1999}]{Kley_1999}
{Kley} W.,  1999, \mndoi [\mnras] {10.1046/j.1365-8711.1999.02198.x}, \href
  {https://ui.adsabs.harvard.edu/abs/1999MNRAS.303..696K} {303, 696}

\bibitem[\protect\citeauthoryear{{Kley} \& {Nelson}}{{Kley} \&
  {Nelson}}{2012}]{KleyNelson2012}
{Kley} W.,  {Nelson} R.~P.,  2012, \mndoi [\araa]
  {10.1146/annurev-astro-081811-125523}, \href
  {https://ui.adsabs.harvard.edu/abs/2012ARA&A..50..211K} {50, 211}

\bibitem[\protect\citeauthoryear{{Kley}, {D'Angelo}  \& {Henning}}{{Kley}
  et~al.}{2001}]{Kley_ea_2001}
{Kley} W.,  {D'Angelo} G.,   {Henning} T.,  2001, \mndoi [\apj]
  {10.1086/318345}, \href
  {https://ui.adsabs.harvard.edu/abs/2001ApJ...547..457K} {547, 457}

\bibitem[\protect\citeauthoryear{{Krajnovi{\'c}}, {Cappellari}, {de Zeeuw}  \&
  {Copin}}{{Krajnovi{\'c}} et~al.}{2006}]{Krajnovic_ea_2006}
{Krajnovi{\'c}} D.,  {Cappellari} M.,  {de Zeeuw} P.~T.,   {Copin} Y.,  2006,
  \mndoi [\mnras] {10.1111/j.1365-2966.2005.09902.x}, \href
  {https://ui.adsabs.harvard.edu/abs/2006MNRAS.366..787K} {366, 787}

\bibitem[\protect\citeauthoryear{{Kratter} \& {Lodato}}{{Kratter} \&
  {Lodato}}{2016}]{Kratter2016}
{Kratter} K.,  {Lodato} G.,  2016, \mndoi [\araa]
  {10.1146/annurev-astro-081915-023307}, \href
  {https://ui.adsabs.harvard.edu/abs/2016ARA&A..54..271K} {54, 271}

\bibitem[\protect\citeauthoryear{{Krijt}, {Schwarz}, {Bergin}  \&
  {Ciesla}}{{Krijt} et~al.}{2018}]{Krijt2018}
{Krijt} S.,  {Schwarz} K.~R.,  {Bergin} E.~A.,   {Ciesla} F.~J.,  2018, \mndoi
  [\apj] {10.3847/1538-4357/aad69b}, \href
  {https://ui.adsabs.harvard.edu/abs/2018ApJ...864...78K} {864, 78}

\bibitem[\protect\citeauthoryear{{Lambrechts} \& {Johansen}}{{Lambrechts} \&
  {Johansen}}{2012}]{LambrechtsJohansen2012}
{Lambrechts} M.,  {Johansen} A.,  2012, \mndoi [\aap]
  {10.1051/0004-6361/201219127}, \href
  {https://ui.adsabs.harvard.edu/abs/2012A&A...544A..32L} {544, A32}

\bibitem[\protect\citeauthoryear{{Le Gal}, {Brady}, {{\"O}berg}, {Roueff}  \&
  {Le Petit}}{{Le Gal} et~al.}{2019}]{LeGal2019}
{Le Gal} R.,  {Brady} M.~T.,  {{\"O}berg} K.~I.,  {Roueff} E.,   {Le Petit} F.,
   2019, \mndoi [\apj] {10.3847/1538-4357/ab4ad9}, \href
  {https://ui.adsabs.harvard.edu/abs/2019ApJ...886...86L} {886, 86}

\bibitem[\protect\citeauthoryear{{Lin} \& {Papaloizou}}{{Lin} \&
  {Papaloizou}}{1986}]{Lin1986}
{Lin} D.~N.~C.,  {Papaloizou} J.,  1986, \mndoi [\apj] {10.1086/164653}, \href
  {https://ui.adsabs.harvard.edu/abs/1986ApJ...309..846L} {309, 846}

\bibitem[\protect\citeauthoryear{{Lin} \& {Youdin}}{{Lin} \&
  {Youdin}}{2015}]{Lin_Youdin_2015}
{Lin} M.-K.,  {Youdin} A.~N.,  2015, \mndoi [\apj]
  {10.1088/0004-637X/811/1/17}, \href
  {https://ui.adsabs.harvard.edu/abs/2015ApJ...811...17L} {811, 17}

\bibitem[\protect\citeauthoryear{{Lodato} et~al.,}{{Lodato}
  et~al.}{2019}]{lodato_2019}
{Lodato} G.,  et~al., 2019, \mndoi [\mnras] {10.1093/mnras/stz913}, \href
  {https://ui.adsabs.harvard.edu/abs/2019MNRAS.486..453L} {486, 453}

\bibitem[\protect\citeauthoryear{{Long} et~al.,}{{Long}
  et~al.}{2018}]{Long_ea_2018}
{Long} F.,  et~al., 2018, \mndoi [\apj] {10.3847/1538-4357/aae8e1}, \href
  {https://ui.adsabs.harvard.edu/abs/2018ApJ...869...17L} {869, 17}

\bibitem[\protect\citeauthoryear{{Lovelace}, {Li}, {Colgate}  \&
  {Nelson}}{{Lovelace} et~al.}{1999}]{lovelace1999}
{Lovelace} R.~V.~E.,  {Li} H.,  {Colgate} S.~A.,   {Nelson} A.~F.,  1999,
  \mndoi [\apj] {10.1086/306900}, \href
  {https://ui.adsabs.harvard.edu/abs/1999ApJ...513..805L} {513, 805}

\bibitem[\protect\citeauthoryear{{Lubow}, {Seibert}  \& {Artymowicz}}{{Lubow}
  et~al.}{1999}]{Lubow_ea_1999}
{Lubow} S.~H.,  {Seibert} M.,   {Artymowicz} P.,  1999, \mndoi [\apj]
  {10.1086/308045}, \href
  {https://ui.adsabs.harvard.edu/abs/1999ApJ...526.1001L} {526, 1001}

\bibitem[\protect\citeauthoryear{{Lyra} \& {Umurhan}}{{Lyra} \&
  {Umurhan}}{2019}]{Lyra_Umurhan2019}
{Lyra} W.,  {Umurhan} O.~M.,  2019, \mndoi [\pasp] {10.1088/1538-3873/aaf5ff},
  \href {https://ui.adsabs.harvard.edu/abs/2019PASP..131g2001L} {131, 072001}

\bibitem[\protect\citeauthoryear{{Lyra}, {Johansen}, {Klahr}  \&
  {Piskunov}}{{Lyra} et~al.}{2008}]{lyra2008}
{Lyra} W.,  {Johansen} A.,  {Klahr} H.,   {Piskunov} N.,  2008, \mndoi [\aap]
  {10.1051/0004-6361:20077948}, \href
  {https://ui.adsabs.harvard.edu/abs/2008A&A...479..883L} {479, 883}

\bibitem[\protect\citeauthoryear{{Lyra}, {Johansen}, {Zsom}, {Klahr}  \&
  {Piskunov}}{{Lyra} et~al.}{2009}]{lyra2009}
{Lyra} W.,  {Johansen} A.,  {Zsom} A.,  {Klahr} H.,   {Piskunov} N.,  2009,
  \mndoi [\aap] {10.1051/0004-6361/200811265}, \href
  {https://ui.adsabs.harvard.edu/abs/2009A&A...497..869L} {497, 869}

\bibitem[\protect\citeauthoryear{{Lyra}, {Turner}  \& {McNally}}{{Lyra}
  et~al.}{2015}]{lyra2015}
{Lyra} W.,  {Turner} N.~J.,   {McNally} C.~P.,  2015, \mndoi [\aap]
  {10.1051/0004-6361/201424919}, \href
  {https://ui.adsabs.harvard.edu/abs/2015A&A...574A..10L} {574, A10}

\bibitem[\protect\citeauthoryear{{Madhusudhan}}{{Madhusudhan}}{2019}]{Madhusudhan2019}
{Madhusudhan} N.,  2019, \mndoi [\araa] {10.1146/annurev-astro-081817-051846},
  \href {https://ui.adsabs.harvard.edu/abs/2019ARA&A..57..617M} {57, 617}

\bibitem[\protect\citeauthoryear{{Marois}, {Macintosh}, {Barman}, {Zuckerman},
  {Song}, {Patience}, {Lafreni{\`e}re}  \& {Doyon}}{{Marois}
  et~al.}{2008}]{marois2008}
{Marois} C.,  {Macintosh} B.,  {Barman} T.,  {Zuckerman} B.,  {Song} I.,
  {Patience} J.,  {Lafreni{\`e}re} D.,   {Doyon} R.,  2008, \mndoi [Science]
  {10.1126/science.1166585}, \href
  {https://ui.adsabs.harvard.edu/abs/2008Sci...322.1348M} {322, 1348}

\bibitem[\protect\citeauthoryear{{Mayor} \& {Queloz}}{{Mayor} \&
  {Queloz}}{1995}]{Mayor_Queloz_1995}
{Mayor} M.,  {Queloz} D.,  1995, \mndoi [\nat] {10.1038/378355a0}, \href
  {https://ui.adsabs.harvard.edu/abs/1995Natur.378..355M} {378, 355}

\bibitem[\protect\citeauthoryear{{Meheut}, {Casse}, {Varniere}  \&
  {Tagger}}{{Meheut} et~al.}{2010}]{Meheut_ea_2010}
{Meheut} H.,  {Casse} F.,  {Varniere} P.,   {Tagger} M.,  2010, \mndoi [\aap]
  {10.1051/0004-6361/201014000}, \href
  {https://ui.adsabs.harvard.edu/abs/2010A&A...516A..31M} {516, A31}

\bibitem[\protect\citeauthoryear{{Meheut}, {Yu}  \& {Lai}}{{Meheut}
  et~al.}{2012a}]{Meheut_ea_2012b}
{Meheut} H.,  {Yu} C.,   {Lai} D.,  2012a, \mndoi [\mnras]
  {10.1111/j.1365-2966.2012.20789.x}, \href
  {https://ui.adsabs.harvard.edu/abs/2012MNRAS.422.2399M} {422, 2399}

\bibitem[\protect\citeauthoryear{{Meheut}, {Meliani}, {Varniere}  \&
  {Benz}}{{Meheut} et~al.}{2012b}]{Meheut_ea_2012a}
{Meheut} H.,  {Meliani} Z.,  {Varniere} P.,   {Benz} W.,  2012b, \mndoi [\aap]
  {10.1051/0004-6361/201219794}, \href
  {https://ui.adsabs.harvard.edu/abs/2012A&A...545A.134M} {545, A134}

\bibitem[\protect\citeauthoryear{{Meru}, {Rosotti}, {Booth}, {Nazari}  \&
  {Clarke}}{{Meru} et~al.}{2019}]{Meru2019}
{Meru} F.,  {Rosotti} G.~P.,  {Booth} R.~A.,  {Nazari} P.,   {Clarke} C.~J.,
  2019, \mndoi [\mnras] {10.1093/mnras/sty2847}, \href
  {https://ui.adsabs.harvard.edu/abs/2019MNRAS.482.3678M} {482, 3678}

\bibitem[\protect\citeauthoryear{{Miotello} et~al.,}{{Miotello}
  et~al.}{2019}]{Miotello2019}
{Miotello} A.,  et~al., 2019, \mndoi [\aap] {10.1051/0004-6361/201935441},
  \href {https://ui.adsabs.harvard.edu/abs/2019A&A...631A..69M} {631, A69}

\bibitem[\protect\citeauthoryear{{Miranda} \& {Rafikov}}{{Miranda} \&
  {Rafikov}}{2019}]{Miranda_Rafikov_2019}
{Miranda} R.,  {Rafikov} R.~R.,  2019, \mndoi [\apjl]
  {10.3847/2041-8213/ab22a7}, \href
  {https://ui.adsabs.harvard.edu/abs/2019ApJ...878L...9M} {878, L9}

\bibitem[\protect\citeauthoryear{{Morbidelli}, {Szul{\'a}gyi}, {Crida}, {Lega},
  {Bitsch}, {Tanigawa}  \& {Kanagawa}}{{Morbidelli}
  et~al.}{2014}]{Morbidelli_ea_2014}
{Morbidelli} A.,  {Szul{\'a}gyi} J.,  {Crida} A.,  {Lega} E.,  {Bitsch} B.,
  {Tanigawa} T.,   {Kanagawa} K.,  2014, \mndoi [Icarus]
  {10.1016/j.icarus.2014.01.010}, \href
  {https://ui.adsabs.harvard.edu/abs/2014Icar..232..266M} {232, 266}

\bibitem[\protect\citeauthoryear{{Mulders}, {Pascucci}, {Apai}  \&
  {Ciesla}}{{Mulders} et~al.}{2018}]{Mulders2018}
{Mulders} G.~D.,  {Pascucci} I.,  {Apai} D.,   {Ciesla} F.~J.,  2018, \mndoi
  [\aj] {10.3847/1538-3881/aac5ea}, \href
  {https://ui.adsabs.harvard.edu/abs/2018AJ....156...24M} {156, 24}

\bibitem[\protect\citeauthoryear{{M{\"u}ller} et~al.,}{{M{\"u}ller}
  et~al.}{2018}]{Mueller_ea_2018}
{M{\"u}ller} A.,  et~al., 2018, \mndoi [\aap] {10.1051/0004-6361/201833584},
  \href {https://ui.adsabs.harvard.edu/abs/2018A&A...617L...2M} {617, L2}

\bibitem[\protect\citeauthoryear{{Nakazato}, {Ikeda}, {Akiyama}, {Kosugi},
  {Yamaguchi}  \& {Honma}}{{Nakazato} et~al.}{2019}]{Nakazato_ea_2019}
{Nakazato} T.,  {Ikeda} S.,  {Akiyama} K.,  {Kosugi} G.,  {Yamaguchi} M.,
  {Honma} M.,  2019, in {Teuben} P.~J.,  {Pound} M.~W.,  {Thomas} B.~A.,
  {Warner} E.~M.,  eds,  Astronomical Society of the Pacific Conference Series
  Vol. 523, Astronomical Data Analysis Software and Systems XXVII. p.~143

\bibitem[\protect\citeauthoryear{{Narayan} \& {Nityananda}}{{Narayan} \&
  {Nityananda}}{1986}]{narayan1986}
{Narayan} R.,  {Nityananda} R.,  1986, \mndoi [\araa]
  {10.1146/annurev.aa.24.090186.001015}, \href
  {https://ui.adsabs.harvard.edu/abs/1986ARA&A..24..127N} {24, 127}

\bibitem[\protect\citeauthoryear{{Nazari}, {Booth}, {Clarke}, {Rosotti},
  {Tazzari}, {Juhasz}  \& {Meru}}{{Nazari} et~al.}{2019}]{Nazari2019}
{Nazari} P.,  {Booth} R.~A.,  {Clarke} C.~J.,  {Rosotti} G.~P.,  {Tazzari} M.,
  {Juhasz} A.,   {Meru} F.,  2019, \mndoi [\mnras] {10.1093/mnras/stz836},
  \href {https://ui.adsabs.harvard.edu/abs/2019MNRAS.485.5914N} {485, 5914}

\bibitem[\protect\citeauthoryear{{Nelson}, {Gressel}  \& {Umurhan}}{{Nelson}
  et~al.}{2013}]{Nelson_ea_2013}
{Nelson} R.~P.,  {Gressel} O.,   {Umurhan} O.~M.,  2013, \mndoi [\mnras]
  {10.1093/mnras/stt1475}, \href
  {https://ui.adsabs.harvard.edu/abs/2013MNRAS.435.2610N} {435, 2610}

\bibitem[\protect\citeauthoryear{{{\"O}berg} \& {Bergin}}{{{\"O}berg} \&
  {Bergin}}{2016}]{Oberg_ea_2016}
{{\"O}berg} K.~I.,  {Bergin} E.~A.,  2016, \mndoi [\apjl]
  {10.3847/2041-8205/831/2/L19}, \href
  {https://ui.adsabs.harvard.edu/abs/2016ApJ...831L..19O} {831, L19}

\bibitem[\protect\citeauthoryear{{{\"O}berg} \& {Wordsworth}}{{{\"O}berg} \&
  {Wordsworth}}{2019}]{Oberg2019}
{{\"O}berg} K.~I.,  {Wordsworth} R.,  2019, \mndoi [\aj]
  {10.3847/1538-3881/ab46a8}, \href
  {https://ui.adsabs.harvard.edu/abs/2019AJ....158..194O} {158, 194}

\bibitem[\protect\citeauthoryear{{{\"O}berg}, {Murray-Clay}  \&
  {Bergin}}{{{\"O}berg} et~al.}{2011}]{Oberg2011}
{{\"O}berg} K.~I.,  {Murray-Clay} R.,   {Bergin} E.~A.,  2011, \mndoi [\apjl]
  {10.1088/2041-8205/743/1/L16}, \href
  {https://ui.adsabs.harvard.edu/abs/2011ApJ...743L..16O} {743, L16}

\bibitem[\protect\citeauthoryear{{{\"O}berg}, {Furuya}, {Loomis}, {Aikawa},
  {Andrews}, {Qi}, {van Dishoeck}  \& {Wilner}}{{{\"O}berg}
  et~al.}{2015}]{Oberg_ea_2015}
{{\"O}berg} K.~I.,  {Furuya} K.,  {Loomis} R.,  {Aikawa} Y.,  {Andrews} S.~M.,
  {Qi} C.,  {van Dishoeck} E.~F.,   {Wilner} D.~J.,  2015, \mndoi [\apj]
  {10.1088/0004-637X/810/2/112}, \href
  {https://ui.adsabs.harvard.edu/abs/2015ApJ...810..112O} {810, 112}

\bibitem[\protect\citeauthoryear{{Ogilvie} \& {Lubow}}{{Ogilvie} \&
  {Lubow}}{2002}]{Ogilvie_Lubow2002}
{Ogilvie} G.~I.,  {Lubow} S.~H.,  2002, \mndoi [\mnras]
  {10.1046/j.1365-8711.2002.05148.x}, \href
  {https://ui.adsabs.harvard.edu/abs/2002MNRAS.330..950O} {330, 950}

\bibitem[\protect\citeauthoryear{{Ormel} \& {Klahr}}{{Ormel} \&
  {Klahr}}{2010}]{OrmelKlahr2010}
{Ormel} C.~W.,  {Klahr} H.~H.,  2010, \mndoi [\aap]
  {10.1051/0004-6361/201014903}, \href
  {https://ui.adsabs.harvard.edu/abs/2010A&A...520A..43O} {520, A43}

\bibitem[\protect\citeauthoryear{{Owen}, {Mahaffy}, {Niemann}, {Atreya},
  {Donahue}, {Bar-Nun}  \& {de Pater}}{{Owen} et~al.}{1999}]{Owen_ea_1999}
{Owen} T.,  {Mahaffy} P.,  {Niemann} H.~B.,  {Atreya} S.,  {Donahue} T.,
  {Bar-Nun} A.,   {de Pater} I.,  1999, \mndoi [\nat] {10.1038/46232}, \href
  {https://ui.adsabs.harvard.edu/abs/1999Natur.402..269O} {402, 269}

\bibitem[\protect\citeauthoryear{{Paardekooper} \& {Mellema}}{{Paardekooper} \&
  {Mellema}}{2004}]{Paardekooper_Mellema2004}
{Paardekooper} S.~J.,  {Mellema} G.,  2004, \mndoi [\aap]
  {10.1051/0004-6361:200400053}, \href
  {https://ui.adsabs.harvard.edu/abs/2004A&A...425L...9P} {425, L9}

\bibitem[\protect\citeauthoryear{{Perez}, {Dunhill}, {Casassus}, {Roman},
  {Szul{\'a}gyi}, {Flores}, {Marino}  \& {Montesinos}}{{Perez}
  et~al.}{2015}]{Perez2015}
{Perez} S.,  {Dunhill} A.,  {Casassus} S.,  {Roman} P.,  {Szul{\'a}gyi} J.,
  {Flores} C.,  {Marino} S.,   {Montesinos} M.,  2015, \mndoi [\apjl]
  {10.1088/2041-8205/811/1/L5}, \href
  {https://ui.adsabs.harvard.edu/abs/2015ApJ...811L...5P} {811, L5}

\bibitem[\protect\citeauthoryear{{P{\'e}rez}, {Casassus}  \&
  {Ben{\'\i}tez-Llambay}}{{P{\'e}rez} et~al.}{2018}]{Perez_ea_2018}
{P{\'e}rez} S.,  {Casassus} S.,   {Ben{\'\i}tez-Llambay} P.,  2018, \mndoi
  [\mnras] {10.1093/mnrasl/sly109}, \href
  {https://ui.adsabs.harvard.edu/abs/2018MNRAS.480L..12P} {480, L12}

\bibitem[\protect\citeauthoryear{{P{\'e}rez}, {Casassus}, {Baruteau}, {Dong},
  {Hales}  \& {Cieza}}{{P{\'e}rez} et~al.}{2019}]{Perez_ea_2019a}
{P{\'e}rez} S.,  {Casassus} S.,  {Baruteau} C.,  {Dong} R.,  {Hales} A.,
  {Cieza} L.,  2019, \mndoi [\aj] {10.3847/1538-3881/ab1f88}, \href
  {https://ui.adsabs.harvard.edu/abs/2019AJ....158...15P} {158, 15}

\bibitem[\protect\citeauthoryear{{P{\'e}rez} et~al.,}{{P{\'e}rez}
  et~al.}{2020}]{Perez_ea_2020}
{P{\'e}rez} S.,  et~al., 2020, \mndoi [\apjl] {10.3847/2041-8213/ab6b2b}, \href
  {https://ui.adsabs.harvard.edu/abs/2020ApJ...889L..24P} {889, L24}

\bibitem[\protect\citeauthoryear{{Pi{\'e}tu}, {Dutrey}  \&
  {Guilloteau}}{{Pi{\'e}tu} et~al.}{2007}]{Pietu_ea_2007}
{Pi{\'e}tu} V.,  {Dutrey} A.,   {Guilloteau} S.,  2007, \mndoi [\aap]
  {10.1051/0004-6361:20066537}, \href
  {https://ui.adsabs.harvard.edu/abs/2007A&A...467..163P} {467, 163}

\bibitem[\protect\citeauthoryear{{Pinte}, {Dent}, {M{\'e}nard}, {Hales},
  {Hill}, {Cortes}  \& {de Gregorio-Monsalvo}}{{Pinte}
  et~al.}{2016}]{Pinte2016}
{Pinte} C.,  {Dent} W.~R.~F.,  {M{\'e}nard} F.,  {Hales} A.,  {Hill} T.,
  {Cortes} P.,   {de Gregorio-Monsalvo} I.,  2016, \mndoi [\apj]
  {10.3847/0004-637X/816/1/25}, \href
  {https://ui.adsabs.harvard.edu/abs/2016ApJ...816...25P} {816, 25}

\bibitem[\protect\citeauthoryear{{Pinte} et~al.,}{{Pinte}
  et~al.}{2018a}]{Pinte_ea_2018a}
{Pinte} C.,  et~al., 2018a, \mndoi [\aap] {10.1051/0004-6361/201731377}, \href
  {https://ui.adsabs.harvard.edu/abs/2018A&A...609A..47P} {609, A47}

\bibitem[\protect\citeauthoryear{{Pinte} et~al.,}{{Pinte}
  et~al.}{2018b}]{Pinte_ea_2018b}
{Pinte} C.,  et~al., 2018b, \mndoi [\apjl] {10.3847/2041-8213/aac6dc}, \href
  {https://ui.adsabs.harvard.edu/abs/2018ApJ...860L..13P} {860, L13}

\bibitem[\protect\citeauthoryear{{Pinte} et~al.,}{{Pinte}
  et~al.}{2019}]{Pinte_ea_2019}
{Pinte} C.,  et~al., 2019, \mndoi [Nature Astronomy]
  {10.1038/s41550-019-0852-6}, \href
  {https://ui.adsabs.harvard.edu/abs/2019NatAs.tmp..419P} {p.~419}

\bibitem[\protect\citeauthoryear{{Pinte} et~al.,}{{Pinte}
  et~al.}{2020}]{Pinte_ea_2020}
{Pinte} C.,  et~al., 2020, arXiv e-prints, \href
  {https://ui.adsabs.harvard.edu/abs/2020arXiv200107720P} {p. arXiv:2001.07720}

\bibitem[\protect\citeauthoryear{{Rafikov}}{{Rafikov}}{2002}]{Rafikov2002}
{Rafikov} R.~R.,  2002, \mndoi [\apj] {10.1086/339399}, \href
  {https://ui.adsabs.harvard.edu/abs/2002ApJ...569..997R} {569, 997}

\bibitem[\protect\citeauthoryear{{Rafikov}}{{Rafikov}}{2009}]{Rafikov2009}
{Rafikov} R.~R.,  2009, \mndoi [\apj] {10.1088/0004-637X/704/1/281}, \href
  {https://ui.adsabs.harvard.edu/abs/2009ApJ...704..281R} {704, 281}

\bibitem[\protect\citeauthoryear{{Riols} \& {Lesur}}{{Riols} \&
  {Lesur}}{2019}]{riols2019}
{Riols} A.,  {Lesur} G.,  2019, \mndoi [\aap] {10.1051/0004-6361/201834813},
  \href {https://ui.adsabs.harvard.edu/abs/2019A&A...625A.108R} {625, A108}

\bibitem[\protect\citeauthoryear{{Riols}, {Lesur}  \& {Menard}}{{Riols}
  et~al.}{2020}]{Riols_ea_2020}
{Riols} A.,  {Lesur} G.,   {Menard} F.,  2020, \mndoi [\aap]
  {10.1051/0004-6361/201937418}, \href
  {https://ui.adsabs.harvard.edu/abs/2020A&A...639A..95R} {639, A95}

\bibitem[\protect\citeauthoryear{{Rosenfeld}, {Andrews}, {Hughes}, {Wilner}  \&
  {Qi}}{{Rosenfeld} et~al.}{2013}]{Rosenfeld_ea_2013}
{Rosenfeld} K.~A.,  {Andrews} S.~M.,  {Hughes} A.~M.,  {Wilner} D.~J.,   {Qi}
  C.,  2013, \mndoi [\apj] {10.1088/0004-637X/774/1/16}, \href
  {https://ui.adsabs.harvard.edu/abs/2013ApJ...774...16R} {774, 16}

\bibitem[\protect\citeauthoryear{{Rosotti}, {Teague}, {Dullemond}, {Booth}  \&
  {Clarke}}{{Rosotti} et~al.}{2020}]{Rosotti_ea_2020}
{Rosotti} G.~P.,  {Teague} R.,  {Dullemond} C.,  {Booth} R.~A.,   {Clarke}
  C.~J.,  2020, \mndoi [\mnras] {10.1093/mnras/staa1170}, \href
  {https://ui.adsabs.harvard.edu/abs/2020MNRAS.495..173R} {495, 173}

\bibitem[\protect\citeauthoryear{{Safronov}}{{Safronov}}{1972}]{Safronov1972}
{Safronov} V.~S.,  1972, {Evolution of the protoplanetary cloud and formation
  of the earth and planets.}

\bibitem[\protect\citeauthoryear{{Schwab}}{{Schwab}}{1984}]{schwab1984}
{Schwab} F.~R.,  1984, in {Roberts} J.~A.,  ed., Indirect Imaging. Measurement
  and Processing for Indirect Imaging. pp 333--346

\bibitem[\protect\citeauthoryear{{Simon} \& {Armitage}}{{Simon} \&
  {Armitage}}{2014}]{simon2014}
{Simon} J.~B.,  {Armitage} P.~J.,  2014, \mndoi [\apj]
  {10.1088/0004-637X/784/1/15}, \href
  {https://ui.adsabs.harvard.edu/abs/2014ApJ...784...15S} {784, 15}

\bibitem[\protect\citeauthoryear{{Simon}, {Beckwith}  \& {Armitage}}{{Simon}
  et~al.}{2012}]{simon2012}
{Simon} J.~B.,  {Beckwith} K.,   {Armitage} P.~J.,  2012, \mndoi [\mnras]
  {10.1111/j.1365-2966.2012.20835.x}, \href
  {https://ui.adsabs.harvard.edu/abs/2012MNRAS.422.2685S} {422, 2685}

\bibitem[\protect\citeauthoryear{{Simon}, {Lesur}, {Kunz}  \&
  {Armitage}}{{Simon} et~al.}{2015}]{simon2015}
{Simon} J.~B.,  {Lesur} G.,  {Kunz} M.~W.,   {Armitage} P.~J.,  2015, \mndoi
  [\mnras] {10.1093/mnras/stv2070}, \href
  {https://ui.adsabs.harvard.edu/abs/2015MNRAS.454.1117S} {454, 1117}

\bibitem[\protect\citeauthoryear{{Simon}, {Bai}, {Flaherty}  \&
  {Hughes}}{{Simon} et~al.}{2018}]{simon2018}
{Simon} J.~B.,  {Bai} X.-N.,  {Flaherty} K.~M.,   {Hughes} A.~M.,  2018, \mndoi
  [\apj] {10.3847/1538-4357/aad86d}, \href
  {https://ui.adsabs.harvard.edu/abs/2018ApJ...865...10S} {865, 10}

\bibitem[\protect\citeauthoryear{{Suriano}, {Li}, {Krasnopolsky}  \&
  {Shang}}{{Suriano} et~al.}{2017}]{suriano2017}
{Suriano} S.~S.,  {Li} Z.-Y.,  {Krasnopolsky} R.,   {Shang} H.,  2017, \mndoi
  [\mnras] {10.1093/mnras/stx735}, \href
  {https://ui.adsabs.harvard.edu/abs/2017MNRAS.468.3850S} {468, 3850}

\bibitem[\protect\citeauthoryear{{Suriano}, {Li}, {Krasnopolsky}  \&
  {Shang}}{{Suriano} et~al.}{2018}]{Suriano_ea_2018}
{Suriano} S.~S.,  {Li} Z.-Y.,  {Krasnopolsky} R.,   {Shang} H.,  2018, \mndoi
  [\mnras] {10.1093/mnras/sty717}, \href
  {https://ui.adsabs.harvard.edu/abs/2018MNRAS.477.1239S} {477, 1239}

\bibitem[\protect\citeauthoryear{{Surville}, {Mayer}  \& {Lin}}{{Surville}
  et~al.}{2016}]{Surville_ea_2016}
{Surville} C.,  {Mayer} L.,   {Lin} D. N.~C.,  2016, \mndoi [\apj]
  {10.3847/0004-637X/831/1/82}, \href
  {https://ui.adsabs.harvard.edu/abs/2016ApJ...831...82S} {831, 82}

\bibitem[\protect\citeauthoryear{{Szul{\'a}gyi} \& {Ercolano}}{{Szul{\'a}gyi}
  \& {Ercolano}}{2020}]{Szulagyi2020}
{Szul{\'a}gyi} J.,  {Ercolano} B.,  2020, arXiv e-prints, \href
  {https://ui.adsabs.harvard.edu/abs/2020arXiv200209918S} {p. arXiv:2002.09918}

\bibitem[\protect\citeauthoryear{{Szul{\'a}gyi}, {Morbidelli}, {Crida}  \&
  {Masset}}{{Szul{\'a}gyi} et~al.}{2014}]{Szulagyi2014}
{Szul{\'a}gyi} J.,  {Morbidelli} A.,  {Crida} A.,   {Masset} F.,  2014, \mndoi
  [\apj] {10.1088/0004-637X/782/2/65}, \href
  {https://ui.adsabs.harvard.edu/abs/2014ApJ...782...65S} {782, 65}

\bibitem[\protect\citeauthoryear{{Szul{\'a}gyi}, {Plas}, {Meyer}, {Pohl},
  {Quanz}, {Mayer}, {Daemgen}  \& {Tamburello}}{{Szul{\'a}gyi}
  et~al.}{2018}]{Szulagyi2018}
{Szul{\'a}gyi} J.,  {Plas} G. v.~d.,  {Meyer} M.~R.,  {Pohl} A.,  {Quanz}
  S.~P.,  {Mayer} L.,  {Daemgen} S.,   {Tamburello} V.,  2018, \mndoi [\mnras]
  {10.1093/mnras/stx2602}, \href
  {https://ui.adsabs.harvard.edu/abs/2018MNRAS.473.3573S} {473, 3573}

\bibitem[\protect\citeauthoryear{{Tanigawa}, {Ohtsuki}  \&
  {Machida}}{{Tanigawa} et~al.}{2012}]{Tanigawa2012}
{Tanigawa} T.,  {Ohtsuki} K.,   {Machida} M.~N.,  2012, \mndoi [\apj]
  {10.1088/0004-637X/747/1/47}, \href
  {https://ui.adsabs.harvard.edu/abs/2012ApJ...747...47T} {747, 47}

\bibitem[\protect\citeauthoryear{{Teague}}{{Teague}}{2019}]{eddy}
{Teague} R.,  2019, \mndoi [The Journal of Open Source Software]
  {10.21105/joss.01220}, \href
  {https://ui.adsabs.harvard.edu/abs/2019JOSS....4.1220T} {4, 1220}

\bibitem[\protect\citeauthoryear{{Teague} \& {Foreman-Mackey}}{{Teague} \&
  {Foreman-Mackey}}{2018}]{Teague_Foreman-Mackey_2018}
{Teague} R.,  {Foreman-Mackey} D.,  2018, \mndoi [Research Notes of the
  American Astronomical Society] {10.3847/2515-5172/aae265}, \href
  {https://ui.adsabs.harvard.edu/abs/2018RNAAS...2..173T} {2, 173}

\bibitem[\protect\citeauthoryear{{Teague} et~al.,}{{Teague}
  et~al.}{2016}]{Teague_ea_2016}
{Teague} R.,  et~al., 2016, \mndoi [\aap] {10.1051/0004-6361/201628550}, \href
  {https://ui.adsabs.harvard.edu/abs/2016A&A...592A..49T} {592, A49}

\bibitem[\protect\citeauthoryear{{Teague} et~al.,}{{Teague}
  et~al.}{2017}]{Teague_ea_2017}
{Teague} R.,  et~al., 2017, \mndoi [\apj] {10.3847/1538-4357/835/2/228}, \href
  {https://ui.adsabs.harvard.edu/abs/2017ApJ...835..228T} {835, 228}

\bibitem[\protect\citeauthoryear{{Teague}, {Bae}, {Bergin}, {Birnstiel}  \&
  {Foreman-Mackey}}{{Teague} et~al.}{2018a}]{Teague_ea_2018a}
{Teague} R.,  {Bae} J.,  {Bergin} E.~A.,  {Birnstiel} T.,   {Foreman-Mackey}
  D.,  2018a, \mndoi [\apjl] {10.3847/2041-8213/aac6d7}, \href
  {https://ui.adsabs.harvard.edu/abs/2018ApJ...860L..12T} {860, L12}

\bibitem[\protect\citeauthoryear{{Teague}, {Bae}, {Birnstiel}  \&
  {Bergin}}{{Teague} et~al.}{2018b}]{Teague_ea_2018b}
{Teague} R.,  {Bae} J.,  {Birnstiel} T.,   {Bergin} E.~A.,  2018b, \mndoi
  [\apj] {10.3847/1538-4357/aae836}, \href
  {https://ui.adsabs.harvard.edu/abs/2018ApJ...868..113T} {868, 113}

\bibitem[\protect\citeauthoryear{{Teague}, {Bae}  \& {Bergin}}{{Teague}
  et~al.}{2019a}]{Teague_ea_2019b}
{Teague} R.,  {Bae} J.,   {Bergin} E.~A.,  2019a, \mndoi [\nat]
  {10.1038/s41586-019-1642-0}, \href
  {https://ui.adsabs.harvard.edu/abs/2019Natur.574..378T} {574, 378}

\bibitem[\protect\citeauthoryear{{Teague}, {Bae}, {Huang}  \&
  {Bergin}}{{Teague} et~al.}{2019b}]{Teague_ea_2019a}
{Teague} R.,  {Bae} J.,  {Huang} J.,   {Bergin} E.~A.,  2019b, \mndoi [\apjl]
  {10.3847/2041-8213/ab4a83}, \href
  {https://ui.adsabs.harvard.edu/abs/2019ApJ...884L..56T} {884, L56}

\bibitem[\protect\citeauthoryear{{Thanathibodee}, {Calvet}, {Bae}, {Muzerolle}
  \& {Hern{\'a}ndez}}{{Thanathibodee} et~al.}{2019}]{Thanathibodee_ea_2019}
{Thanathibodee} T.,  {Calvet} N.,  {Bae} J.,  {Muzerolle} J.,   {Hern{\'a}ndez}
  R.~F.,  2019, \mndoi [\apj] {10.3847/1538-4357/ab44c1}, \href
  {https://ui.adsabs.harvard.edu/abs/2019ApJ...885...94T} {885, 94}

\bibitem[\protect\citeauthoryear{{Thompson}, {Moran}  \& {Swenson}}{{Thompson}
  et~al.}{2017}]{thompson_2017}
{Thompson} A.~R.,  {Moran} J.~M.,   {Swenson} George~W. J.,  2017,
  {Interferometry and Synthesis in Radio Astronomy, 3rd Edition},
  \mndoi{10.1007/978-3-319-44431-4.
}

\bibitem[\protect\citeauthoryear{{Uribe}, {Klahr}, {Flock}  \&
  {Henning}}{{Uribe} et~al.}{2011}]{uribe2011}
{Uribe} A.~L.,  {Klahr} H.,  {Flock} M.,   {Henning} T.,  2011, \mndoi [\apj]
  {10.1088/0004-637X/736/2/85}, \href
  {https://ui.adsabs.harvard.edu/abs/2011ApJ...736...85U} {736, 85}

\bibitem[\protect\citeauthoryear{{Vigan} et~al.,}{{Vigan}
  et~al.}{2017}]{vigan2017}
{Vigan} A.,  et~al., 2017, \mndoi [\aap] {10.1051/0004-6361/201630133}, \href
  {https://ui.adsabs.harvard.edu/abs/2017A&A...603A...3V} {603, A3}

\bibitem[\protect\citeauthoryear{{Vigan} et~al.,}{{Vigan}
  et~al.}{2020}]{Vigan_ea_2020}
{Vigan} A.,  et~al., 2020, arXiv e-prints, \href
  {https://ui.adsabs.harvard.edu/abs/2020arXiv200706573V} {p. arXiv:2007.06573}

\bibitem[\protect\citeauthoryear{{Wagner} et~al.,}{{Wagner}
  et~al.}{2018}]{Wagner2018}
{Wagner} K.,  et~al., 2018, \mndoi [\apjl] {10.3847/2041-8213/aad695}, \href
  {https://ui.adsabs.harvard.edu/abs/2018ApJ...863L...8W} {863, L8}

\bibitem[\protect\citeauthoryear{{Walsh}, {Daley}, {Facchini}  \&
  {Juh{\'a}sz}}{{Walsh} et~al.}{2017}]{Walsh_ea_2017}
{Walsh} C.,  {Daley} C.,  {Facchini} S.,   {Juh{\'a}sz} A.,  2017, \mndoi
  [\aap] {10.1051/0004-6361/201731334}, \href
  {https://ui.adsabs.harvard.edu/abs/2017A&A...607A.114W} {607, A114}

\bibitem[\protect\citeauthoryear{{Walter}, {Brinks}, {de Blok}, {Bigiel},
  {Kennicutt}, {Thornley}  \& {Leroy}}{{Walter} et~al.}{2008}]{Walter_ea_2008}
{Walter} F.,  {Brinks} E.,  {de Blok} W.~J.~G.,  {Bigiel} F.,  {Kennicutt}
  Robert~C. J.,  {Thornley} M.~D.,   {Leroy} A.,  2008, \mndoi [\aj]
  {10.1088/0004-6256/136/6/2563}, \href
  {https://ui.adsabs.harvard.edu/abs/2008AJ....136.2563W} {136, 2563}

\bibitem[\protect\citeauthoryear{{Ward}}{{Ward}}{1976}]{Ward1976}
{Ward} W.~R.,  1976, in {Avrett} E.~H.,  ed., Frontiers of Astrophysics. pp
  1--40

\bibitem[\protect\citeauthoryear{Ward}{Ward}{2000}]{ward2000}
Ward W.~R.,  2000, Origin of the Earth and Moon, pp 75--84

\bibitem[\protect\citeauthoryear{{Whipple}}{{Whipple}}{1972}]{Whipple_1972}
{Whipple} F.~L.,  1972, in {Elvius} A.,  ed., From Plasma to Planet. p.~211

\bibitem[\protect\citeauthoryear{{Winn} \& {Fabrycky}}{{Winn} \&
  {Fabrycky}}{2015}]{winn_fabrycky_2015}
{Winn} J.~N.,  {Fabrycky} D.~C.,  2015, \mndoi [\araa]
  {10.1146/annurev-astro-082214-122246}, \href
  {https://ui.adsabs.harvard.edu/abs/2015ARA&A..53..409W} {53, 409}

\bibitem[\protect\citeauthoryear{{Wootten} \& {Thompson}}{{Wootten} \&
  {Thompson}}{2009}]{wootten2009}
{Wootten} A.,  {Thompson} A.~R.,  2009, \mndoi [IEEE Proceedings]
  {10.1109/JPROC.2009.2020572}, \href
  {https://ui.adsabs.harvard.edu/abs/2009IEEEP..97.1463W} {97, 1463}

\bibitem[\protect\citeauthoryear{{Yen}, {Koch}, {Liu}, {Puspitaningrum},
  {Hirano}, {Lee}  \& {Takakuwa}}{{Yen} et~al.}{2016}]{Yen_ea_2016}
{Yen} H.-W.,  {Koch} P.~M.,  {Liu} H.~B.,  {Puspitaningrum} E.,  {Hirano} N.,
  {Lee} C.-F.,   {Takakuwa} S.,  2016, \mndoi [\apj]
  {10.3847/0004-637X/832/2/204}, \href
  {https://ui.adsabs.harvard.edu/abs/2016ApJ...832..204Y} {832, 204}

\bibitem[\protect\citeauthoryear{Zhang, Blake  \& Bergin}{Zhang
  et~al.}{2015}]{zhang2015}
Zhang K.,  Blake G.~A.,   Bergin E.~A.,  2015, The Astrophysical Journal
  Letters, 806, L7

\bibitem[\protect\citeauthoryear{{Zhang} et~al.,}{{Zhang}
  et~al.}{2018}]{Zhang2018}
{Zhang} S.,  et~al., 2018, \mndoi [\apjl] {10.3847/2041-8213/aaf744}, \href
  {https://ui.adsabs.harvard.edu/abs/2018ApJ...869L..47Z} {869, L47}

\bibitem[\protect\citeauthoryear{{Zhang}, {Bergin}, {Schwarz}, {Krijt}  \&
  {Ciesla}}{{Zhang} et~al.}{2019}]{Zhang2019}
{Zhang} K.,  {Bergin} E.~A.,  {Schwarz} K.,  {Krijt} S.,   {Ciesla} F.,  2019,
  \mndoi [\apj] {10.3847/1538-4357/ab38b9}, \href
  {https://ui.adsabs.harvard.edu/abs/2019ApJ...883...98Z} {883, 98}

\bibitem[\protect\citeauthoryear{{Zhu}, {Stone}  \& {Rafikov}}{{Zhu}
  et~al.}{2012}]{Zhu2012}
{Zhu} Z.,  {Stone} J.~M.,   {Rafikov} R.~R.,  2012, \mndoi [\apjl]
  {10.1088/2041-8205/758/2/L42}, \href
  {https://ui.adsabs.harvard.edu/abs/2012ApJ...758L..42Z} {758, L42}

\bibitem[\protect\citeauthoryear{{Zhu}, {Dong}, {Stone}  \& {Rafikov}}{{Zhu}
  et~al.}{2015}]{Zhu2015}
{Zhu} Z.,  {Dong} R.,  {Stone} J.~M.,   {Rafikov} R.~R.,  2015, \mndoi [\apj]
  {10.1088/0004-637X/813/2/88}, \href
  {https://ui.adsabs.harvard.edu/abs/2015ApJ...813...88Z} {813, 88}

\bibitem[\protect\citeauthoryear{{Zurlo} et~al.,}{{Zurlo}
  et~al.}{2020}]{Zurlo2020}
{Zurlo} A.,  et~al., 2020, \mndoi [\aap] {10.1051/0004-6361/201936891}, \href
  {https://ui.adsabs.harvard.edu/abs/2020A&A...633A.119Z} {633, A119}

\bibitem[\protect\citeauthoryear{{de Gregorio-Monsalvo} et~al.,}{{de
  Gregorio-Monsalvo} et~al.}{2013}]{deGregorio-Monsalvo_ea_2013}
{de Gregorio-Monsalvo} I.,  et~al., 2013, \mndoi [\aap]
  {10.1051/0004-6361/201321603}, \href
  {https://ui.adsabs.harvard.edu/abs/2013A&A...557A.133D} {557, A133}

\bibitem[\protect\citeauthoryear{{van Boekel} et~al.,}{{van Boekel}
  et~al.}{2017}]{vanBoekel_ea_2017}
{van Boekel} R.,  et~al., 2017, \mndoi [\apj] {10.3847/1538-4357/aa5d68}, \href
  {https://ui.adsabs.harvard.edu/abs/2017ApJ...837..132V} {837, 132}

\makeatother
\end{thebibliography}
\bibliographystyle{pasa-mnras}

\end{document}